\documentclass[pra,aps,showpacs,twocolumn]{revtex4-1}

\usepackage[latin1]{inputenc}
\usepackage{amsmath,amssymb,amscd,bbm,epsfig}
\usepackage[all,cmtip]{xy}
\usepackage{color}

\newcommand{\mathd}{\mathrm{d}}
\newcommand{\mathe}{\mathrm{e}}

\newcommand{\tmmathbf}[1]{\ensuremath{\boldsymbol{#1}}}
\newcommand{\tmop}[1]{\ensuremath{\operatorname{#1}}}
\newcommand{\tmtextbf}[1]{{\bfseries{#1}}}
\newcommand{\tmtextit}[1]{{\itshape{#1}}}

\begin{document}

\title{Markovian and non-Markovian dynamics in quantum and classical systems}

\author{Bassano \surname{Vacchini}}

\email{bassano.vacchini@mi.infn.it}

\author{Andrea \surname{Smirne}}

\affiliation{Dipartimento di Fisica, Universit{\`a} degli Studi di
Milano, Via Celoria 16, I-20133 Milan, Italy}

\affiliation{INFN, Sezione di Milano, Via Celoria 16, I-20133
Milan, Italy}

\author{Elsi-Mari Laine}

\author{Jyrki Piilo}

\affiliation{Turku Centre for Quantum Physics, Department of
Physics and Astronomy, University of Turku, FI-20014 Turun
yliopisto, Finland}

\author{Heinz-Peter Breuer}

\affiliation{Physikalisches Institut, Universit\"at Freiburg,
Hermann-Herder-Strasse 3, D-79104 Freiburg, Germany}

\date{\today}

\begin{abstract}
   We discuss the conceptually different definitions used for the
   non-Markovianity of classical and quantum processes. The
   well-established definition for non-Markovianity of a classical
   stochastic process represents a condition on the Kolmogorov
   hierarchy of the $n$-point joint probability distributions. Since
   this definition cannot be transferred to the quantum regime,
   quantum non-Markovianity has recently been defined and quantified
   in terms of the underlying quantum dynamical map, using either its
   divisibility properties or the behavior of the trace distance
   between pairs of initial states. Here, we investigate and compare
   these definitions and their relations to the classical notion of
   non-Markovianity by employing a large class of non-Markovian
   processes, known as semi-Markov processes, which admit a natural
   extension to the quantum case. A number of specific physical
   examples is constructed which allow to study the basic features of
   the classical and the quantum definitions and to evaluate
   explicitly the measures for quantum non-Markovianity. Our results
   clearly demonstrate several fundamental distinctions between the
   classical and the quantum notion of non-Markovianity, as well as
   between the various quantum measures for non-Markovianity.
\end{abstract}


\maketitle

\section{Introduction\label{sec:intro}}

A lot of effort has been put into the study and understanding of non-Markovian
effects and dynamics within the description of open quantum systems
{\cite{Breuer2002}}. These efforts have faced quite a huge amount of
difficulties arising both because of practical as well as fundamental reasons.
Indeed, on the one hand the treatment of non-Markovian systems is especially
demanding because one cannot rely on simplifying assumptions such as weak
coupling, separation of time scales between system and environment, and the
factorization of the system-environment state. On the other hand, in the
non-Markovian case there is no general characterization of the equations of
motion, such as in the Markovian setting thanks to the theory of quantum
dynamical semigroups, and the very notion of non-Markovianity in the quantum
case still has to be cleared up. In the last few years this topic has
experienced a significant revival, leading to important improvements and to the
deeper understanding of quite a few issues in the theory of open quantum
systems (see e.g.
{\cite{Stockburger2002a,Daffer2004a,Piilo2008a,Breuer2008a,Shabani2009a,Chruscinski2010a,Barchielli2010a,Laine2010c}}
and references therein).

Apart from the explicit detailed treatment of many specific quantum systems
where memory effects show up, efforts have been made to obtain general classes
of non-Markovian equations leading to well-defined completely positive time
evolutions {\cite{Budini2004a,Breuer2008a}}. At the same time advances have
been made in order to actually define what is meant by a non-Markovian quantum
dynamics {\cite{Wolf2008a,Breuer2009b,Rivas2010a}}. This work has raised all
sorts of questions regarding the connection among the different approaches, as
well as the relationship between the notion of non-Markovianity used in
classical and quantum setting, together with the quest for clearcut signatures
of non-Markovian behavior.

The present paper is devoted to address some of these questions, focussing in
particular on the connection between the very definition of non-Markovian
process used in classical probability theory, and the Markovian or
non-Markovian behavior in the dynamics of a physical system. It will naturally
appear that these two notions are quite different. Starting within the
classical framework, we will analyze how the non-Markovianity of a process
reflects itself in the behavior of its one-point probability density, which
naturally leads to criteria for the characterization of non-Markovian behavior
in the dynamics. These criteria can be extended to the quantum setting, thus
providing natural tools to assess the non-Markovianity of a quantum time
evolution. They are based on the possibility to connect the probability
vectors giving the state of the system at different times through well defined
transition matrices, and to the behavior of solutions corresponding to
different initial states with respect to the Kolmogorov distance.

Despite the abstract framework, the whole presentation is built with reference
to explicit examples. These examples find their common root in being related
to realizations of a class of non-Markovian processes for which, as an
exceptional case, an explicit characterization is available, namely
semi-Markov processes. This approach will enable us to construct different non
trivial dynamics both in the classical and the quantum case, showing quite
different behaviors and amenable to an explicit analysis, both in terms of the
previously introduced criteria and of the recently proposed measures of
non-Markovianity, thus allowing for a concrete comparison of these measures.

The paper is organized as follows. In Sect.~\ref{sec:pcpdiv} we introduce the
notion of classical and quantum map for the dynamics of an open system,
introducing the notions of P-divisibility and CP-divisibility which will turn
out to be useful in order to compare different behaviors. In
Sect.~\ref{sec:cnm} we recall the notion of non-Markovianity for a classical
process, showing its relation to the behavior of the one-point probability
density. We further consider examples of classical semi-Markov processes
characterized by different stochastic matrices and waiting time distributions,
studying their P-divisibility and the behavior with respect to the Kolmogorov
distance. In Sect.~\ref{sec:qnm} we perform a similar analysis in the quantum
setting, considering classes of dynamics which can be related to semi-Markov
processes. These dynamics still allow for an exact determination of their
divisibility properties and of their quantum measure of non-Markovianity
according to the recent proposals. In particular we provide exact expressions
for the values of these measures, thus allowing for their explicit comparison.
We further comment on our results in Sect.~\ref{sec:ceo}.

\section{Dynamical maps\label{sec:pcpdiv}}

If the initial state of system and environment factorizes, the dynamics of an
open quantum system can be described by a trace preserving completely positive
(CPT) map, so that the state of the system $\rho \left( t \right)$ evolving
from an initial state $\rho \left( t_0 \right)$ is given by
\begin{eqnarray}
  \rho \left( t \right) & = & \Phi \left( t, t_0 \right) \rho \left( t_0
  \right) .  \label{eq:qmap}
\end{eqnarray}
The most general structure of such time evolution maps is not known, apart
from the important subclass of maps which obey a semigroup composition law,
which can be characterized via a suitable generator
{\cite{Lindblad1976a,Gorini1976a}}. In order to characterize and actually
define Markovianity or non-Markovianity in this setting one can follow
essentially two approaches: either one assumes that the map $\Phi \left( t,
t_0 \right)$ is known, and therefore relys on looking at certain mathematical
properties of the map itself, which is essentially the path followed in
different ways in {\cite{Wolf2008a,Rivas2010a}}; or one studies the behavior
in time of the solutions $\rho \left( t \right)$ allowing the initial
condition to vary over the possible set of states, which is the approach
elaborated in {\cite{Breuer2009b,Laine2010a}} relying on a suitable notion of
distinguishability of quantum states {\cite{Hayashi2006}}, an approach which
captures the idea of information flow between system and environment. Note
that Markovianity or non-Markovianity should be a property of the map or
equivalently of the time evolved states, not of the equations admitting such
states as solutions. Indeed quite different form of the equations, e.g.
integrodifferential or local in time, might admit the very same solutions, as
we shall also see in the examples in Sect.~\ref{sec:qnm}. Of course, since in
a concrete physical setting one is rather faced with the equations of motion
rather than with their general solution, it is therefore also of great interest to assess
possible links between the equations themselves and the Markovian or
non-Markovian behavior of their solutions.

To better understand and compare these two approaches, besides the notion of
complete positivity and trace preservation other finer characterizations of
the time evolution map $\Phi \left( t, 0 \right)$ turn out to be useful, where
we have set for the sake of simplicity $t_0 = 0$. If the map $\Phi \left( t, 0
\right)$ can be split according to
\begin{eqnarray}
  \Phi \left( t, 0 \right) & = & \Phi \left( t, s \right) \Phi \left( s, 0
  \right)  \label{eq:qdiv}
\end{eqnarray}
for any $t \geqslant s \geqslant 0$, with $\Phi \left( t, s \right)$ itself a
CPT map, we say that the map $\Phi \left( t, 0 \right)$ is CP-divisible, which
implies that $\Phi \left( t, s \right)$ is itself a well-defined time
evolution having as domain the whole set of states. Note that at variance with
Ref.~{\cite{Wolf2008b}} focussing on quantum channels, the notion of
divisibility considered here refers to families of time dependent dynamical
maps. The existence of $\Phi \left( t, s \right)$ as a linear map is granted
if $\Phi \left( s, 0 \right)$ is invertible, which is typically the case away
from isolated points of time, so that
\begin{eqnarray}
  \Phi \left( t, s \right) & = & \Phi \left( t, 0 \right) \Phi^{- 1} \left( s,
  0 \right) .  \label{eq:trick}
\end{eqnarray}
The existence of $\Phi \left( t, s \right)$ as a linear map, however, does not
entail its complete positivity. Indeed, we will say that the map $\Phi \left(
t, 0 \right)$ is P-divisible if $\Phi \left( t, s \right)$ send states into
states but is only positive, and that it is indivisible if neither
P-divisibility nor CP-divisibility hold. Examples of such maps together with
their physical interpretation will be provided in Sect.~\ref{sec:qnm}.

The notion of P-divisibility can be considered also in the classical setting.
Suppose to consider a finite dimensional classical system, described by a
probability vector $\tmmathbf{p} \left( t \right)$. Its time evolution can be
described by a time dependent collection of stochastic matrices $\Lambda
\left( t, 0 \right)$ according to
\begin{eqnarray}
  \tmmathbf{p} \left( t \right) & = & \Lambda \left( t, 0 \right) \tmmathbf{p}
  \left( 0 \right) .  \label{eq:cmap}
\end{eqnarray}
Similarly as before, we say that the classical map $\Lambda \left( t, 0
\right)$ is P-divisible provided for any $t \geqslant s \geqslant 0$ one can
write
\begin{eqnarray}
  \Lambda \left( t, 0 \right) & = & \Lambda \left( t, s \right) \Lambda \left(
  s, 0 \right),  \label{eq:cdiv}
\end{eqnarray}
where each of the $\Lambda \left( t, s \right)$ is itself a stochastic matrix.
Its matrix elements satisfy therefore $\left( \Lambda \left( t, s \right)
\right)_{ij} \geqslant 0$ and $\sum_k \left( \Lambda \left( t, s \right)
\right)_{kj} = 1$, which provide the necessary and sufficient conditions
ensuring that probability vectors are sent into probability vectors
{\cite{Norris1999}}. Once again this need not generally be true, even if the
map $\Lambda \left( s, 0 \right)$ is invertible as linear operator. Note that
we are here only considering the one-point probabilities $\tmmathbf{p} \left(
t \right)$, which are certainly not enough to assess Markovianity or
non-Markovianity of a process according to the mathematically precise
definition used in classical probability theory.

\section{Classical non-Markovian processes\label{sec:cnm}}

Let us now recall what is the very definition of non-Markovian process in the
classical probabilistic setting. \ Indeed the analysis of classical processes
is a natural starting point, also adopted in
{\cite{Chruscinski2010c,Chruscinski-xxx2,Chruscinski-xxx3,Rivas-xxx,Karimipour-xxx}}.
Suppose we are considering a stochastic process taking values in a numerable
set $\left\{ x_i \right\}_{i \in \mathbbm{N}}$. The process is said to be
Markovian if the conditional transition probabilities satisfy
\begin{multline}
  p_{1| n} \left( x_n, t_n |x_{n - 1}, t_{n - 1} ; \ldots ; x_0, t_0 \right) 
  = 
\\
 p_{1|1} \left( x_n, t_n |x_{n - 1}, t_{n - 1} \right)  \label{eq:mp}
\end{multline}
with $t_n \geqslant t_{n - 1} \geqslant \ldots \geqslant t_1 \geqslant t_0$,
so that the probability that the random variable assumes the value $x_n$ at
time $t_n$, given that it has assumed given values $x_i$ at previous times
$t_i$, actually only depends on the last assumed value, and not on previous
ones. In this sense the process is said to lack memory. This statement
obviously involves all $n$-times probabilities, so that the non-Markovianity
of the process cannot be assessed by looking at the one-time probabilities
only {\cite{VanKampen1998a,Gillespie1998a}}. It immediately appears from
Eq.~(\ref{eq:mp}) that such a definition cannot be transferred to the quantum
realm, since the very notion of conditional probability does depend on the
measurement performed in order to ascertain the previous value of the random
variable, and on how it transforms the state for the subsequent time
evolution.

If we know that the process is Markovian the $n$-point probabilities can be
easily obtained in terms of the initial probability density and the
conditional transition probability $p_{1|1}$ according to
\begin{multline}
  p_n \left( x_n, t_n ; x_{n - 1}, t_{n - 1} ; \ldots ; x_0, t_0
  \right)  =
\\ 
  \prod_{i = 1}^n p_{1|1} \left( x_i, t_i |x_{i - 1}, t_{i - 1} \right) p_1
  \left( x_0, t_0 \right) .  \label{eq:np}
\end{multline}
The Markov condition Eq.~(\ref{eq:mp}) in turn implies that the conditional
transition probability $p_{1|1}$ obeys the Chapman-Kolmogorov equation
\begin{eqnarray}
  p_{1|1} \left( x, t|y, s \right) & = & \sum_z p_{1|1} \left( x, t|z, \tau
  \right) p_{1|1} \left( z, \tau |y, s \right),  \label{eq:ck}
\end{eqnarray}
with $t \geqslant \tau \geqslant s$, whose possible solutions characterize the
transition probabilities of a Markov process. A Markov process is therefore
uniquely characterized by its conditional transition probability and the
initial distribution.

If $\tmmathbf{p} \left( t \right)$ denotes the vector giving the one-point
probability of a Markov process taking values in a finite space, and if
$\Lambda \left( t, s \right)$ is its conditional transition probability expressed in
matrix form, the probability vectors at different times are related according
to
\begin{eqnarray}
  \tmmathbf{p} \left( t \right) & = & \Lambda \left( t, s \right) \tmmathbf{p}
  \left( s \right)  \label{eq:plp}
\end{eqnarray}
with $t \geqslant s \geqslant 0$ and $\Lambda \left( t, s \right)$ a
stochastic matrix obeying the Chapman-Kolmogorov equation, which in this case is equivalent to the requirement of P-divisibility in the sense of
Eq.~(\ref{eq:cdiv}). However, in general validity of the Chapman-Kolmogorov equation and
P-divisibility obviously do not coincide, since the P-divisibility does not
tell anything about the higher-point conditional probabilities. Indeed, for a
given process one might find a matrix $\Lambda \left( t, s \right)$ satisfying
Eq.~(\ref{eq:plp}) even if the process is non-Markovian, however in this case
the matrix is not the conditional transition probability of the process
{\cite{Hanggi1977a,Hanggi1982a}}. Similarly to Eq.~(\ref{eq:trick}), if
$\Lambda \left( t, 0 \right)$ is invertible one can obtain such a matrix as
\begin{eqnarray}
  \Lambda \left( t, s \right) & = & \Lambda \left( t, 0 \right) \Lambda^{- 1}
  \left( s, 0 \right),  \label{eq:trick2}
\end{eqnarray}
which warrants independence from the initial probability vector. On the
contrary, the conditional transition probability of a non-Markovian process does depend on
the initial probability vector {\cite{Hanggi1977a}}.

\subsection{Classical semi-Markov processes\label{sec:csmp}}

To spell out this situation in detail let us consider an example. The main
difficulty lies in the fact that generally very little is known on
non-Markovian processes. We will however consider a class of non-Markovian
processes allowing for a compact characterization, that is to say semi-Markov
processes {\cite{Feller1971}}. To this end we consider a system having a
finite set of states, which can jump from one state to the other according to
certain jump probabilities $\pi_{mn}$, waiting a random time in each state
before jumping to the next. Such a process turns out to be Markovian if and
only if the waiting time distributions $f_n \left( \tau \right)$
characterizing the random sojourn time spent in each state are given by
exponential probability distributions, and non-Markovian otherwise. Each such process is fixed by its semi-Markov matrix
\begin{eqnarray}
  \left( Q \right)_{mn} \left( \tau \right) & = & \pi_{mn} f_n \left( \tau
  \right),  \label{eq:smm}
\end{eqnarray}
where the $\pi_{mn}$ are the elements of a stochastic matrix. Assuming for the
sake of simplicity a two-dimensional system, a site independent waiting time
distribution and the stochastic matrix to be actually bistochastic
{\cite{Norris1999}}, the semi-Markov matrix is determined according to
\begin{eqnarray}
  Q \left( \tau \right) & = & \left(\begin{array}{cc}
    1 - \pi & \pi\\
    \pi & 1 - \pi
  \end{array}\right) f \left( \tau \right), \nonumber  \label{eq:qbisto}  \\
  & \equiv & \Pi f \left( \tau \right) 
\end{eqnarray}
with $\pi$ a positive number between zero and one giving the probability to
jump from one site to the other, and $f \left( \tau \right)$ an arbitrary
waiting time distribution with associated survival probability
\begin{eqnarray}
  g \left( t \right) & = & 1 - \int_0^t \mathd \tau f \left( \tau \right) . 
  \label{eq:sur}
\end{eqnarray}

\subsection{Transition probability}

The transition probability for such a process can be determined exploiting the
fact that it is known to obey the following integrodifferential Kolmogorov
forward equation {\cite{Feller1964a}}
\begin{eqnarray}
  \frac{\mathd}{\tmop{dt}} T \left( t, 0 \right) & = & \int_0^t \mathd \tau W
  \left( \tau \right) T \left( t - \tau, 0 \right),  \label{eq:fk}
\end{eqnarray}
here expressed in matrix form with
\begin{eqnarray}
  W \left( \tau \right) & = & \left( \Pi - \mathbbm{1} \right) k \left( \tau
  \right),  \label{eq:wmatrix}
\end{eqnarray}
where the memory kernel $k \left( \tau \right)$ relates waiting time
distribution and survival probability according to
\begin{eqnarray}
  f \left( \tau \right) & = & \int_0^{\tau} \mathd tk \left( \tau - t \right)
  g \left( t \right) .  \label{eq:k}
\end{eqnarray}
Denoting by $\hat{v} \left( u \right)$ the Laplace transform of a function or
matrix $v \left( \tau \right)$ according to
\begin{eqnarray}
  \hat{v} \left( u \right) & = & \int^{+ \infty}_0 \mathd \tau v \left( \tau
  \right) \mathe^{- u \tau},  \label{eq:lt}
\end{eqnarray}
the solution of this equation with initial condition $T \left( 0, 0 \right) =
\mathbbm{1}$ can be expressed as
\begin{eqnarray}
  \hat{T} \left( u \right) & = & \frac{\hat{g} \left( u \right)}{\mathbbm{1} -
  \Pi \hat{f} \left( u \right)} .  \label{eq:tu}
\end{eqnarray}
\subsection{Explicit examples}

The explicit solution of Eq.~(\ref{eq:tu}) for $\pi = 1 / 2$, so that at each
step the system has equal probability to remain in the same site or change, is
given by
\begin{eqnarray}
  T \left( t, 0 \right) & = & \frac{1}{2} \left(\begin{array}{cc}
    1 + g \left( t \right) & 1 - g \left( t \right)\\
    1 - g \left( t \right) & 1 + g \left( t \right)
  \end{array}\right),  \label{eq:tpi12}
\end{eqnarray}
so that according to Eq.~(\ref{eq:trick2}) we can introduce the matrices
\begin{eqnarray}
  T \left( t, s \right) & = & T \left( t, 0 \right) T^{- 1} \left( s, 0
  \right) \nonumber\\
  & = & \frac{1}{2} \left(\begin{array}{cc}
    1 + g \left( t \right) / g \left( s \right) & 1 - g \left( t \right) / g
    \left( s \right)\\
    1 - g \left( t \right) / g \left( s \right) & 1 + g \left( t \right) / g
    \left( s \right)
  \end{array}\right),  \label{eq:12ts}
\end{eqnarray}
which indeed connect the probability vectors at different times according to
\begin{eqnarray}
  \tmmathbf{p} \left( t \right) & = & T \left( t, s \right) \tmmathbf{p}
  \left( s \right) .  \label{eq:11}
\end{eqnarray}
Given the fact that for any non vanishing waiting time distribution the
survival probability is a strictly positive monotonic decreasing function, the
matrices $T \left( t, s \right)$ are well-defined stochastic matrices for any
pair of times $t \geqslant s$, so that the classical map $T \left( t, 0
\right)$ is always P-divisible, irrespective of the fact that the underlying
process is Markovian only for the special choice of an exponential waiting
time distribution of the form
\begin{eqnarray}
  f \left( \tau \right) & = & \lambda \mathe^{- \lambda \tau} . 
  \label{eq:expwtd}
\end{eqnarray}
This result implies that the one-point probabilities of the considered
non-Markovian semi-Markov process can be equally well obtained from a Markov
process with conditional transition probability $p_{1|1}$ given by $T$, whose
$n$-point probabilities can be obtained as in Eq.~(\ref{eq:np}). The latter
would however differ from those of the considered semi-Markov process.

As a complementary situation, let us consider the case $\pi = 1$, so that once
in a state the system jumps with certainty to the other, thus obtaining as
explicit solution of Eq.~(\ref{eq:tu}) the expression
\begin{eqnarray}
  T \left( t, 0 \right) & = & \frac{1}{2} \left(\begin{array}{cc}
    1 + q \left( t \right) & 1 - q \left( t \right)\\
    1 - q \left( t \right) & 1 + q \left( t \right)
  \end{array}\right),  \label{eq:tpi1}
\end{eqnarray}
and therefore
\begin{eqnarray}
  T \left( t, s \right) & = & T \left( t, 0 \right) T^{- 1} \left( s, 0
  \right) \nonumber\\
  & = & \frac{1}{2} \left(\begin{array}{cc}
    1 + q \left( t \right) / q \left( s \right) & 1 - q \left( t \right) / q
    \left( s \right)\\
    1 - q \left( t \right) / q \left( s \right) & 1 + q \left( t \right) / q
    \left( s \right)
  \end{array}\right) .  \label{eq:1ts}
\end{eqnarray}
The quantity $q \left( t \right)$ appearing in these matrices is the inverse
Laplace transform of the function
\begin{eqnarray}
  \hat{q} \left( u \right) & = & \frac{1}{u}  \frac{1 - \hat{f} \left( u
  \right)}{1 + \hat{f} \left( u \right)} .  \label{eq:qu}
\end{eqnarray}
Recalling that the probability for $n$ jumps in a time $t$ for a waiting time
distribution $f \left( t \right)$ is given by
\begin{eqnarray}
  p_n \left( t \right) = \int^t_0 \mathd \tau f \left( t - \tau \right) p_{n -
  1} \left( \tau \right), &  &  \label{eq:pn}
\end{eqnarray}
so that
\begin{eqnarray}
  \hat{p}_n \left( u \right) & = & \hat{p}_0 \left( u \right) \hat{f}^n \left(
  u \right),  \label{eq:pnu}
\end{eqnarray}
one has
\begin{eqnarray}
  q \left( t \right) & = & \sum^{\infty}_{n = 0} p_{2 n} (t) -
  \sum^{\infty}_{n = 0} p_{2 n + 1} (t)  \label{eq:qdiff}\\
  & = & p_{\tmop{even}} \left( t \right) - p_{\tmop{odd}} \left( t \right) .
  \nonumber
\end{eqnarray}
The quantity $q \left( t \right)$ therefore expresses the difference between
the probability to have an even or an odd number of jumps. At variance with
the previous situation, the quantity $q \left( t \right)$ depending on the
waiting time distribution can assume quite different behavior, showing
oscillations and going through zero at isolated time points, so that at these
time points, corresponding to a crossing of trajectories starting from
different initial conditions, the transition matrix $T \left( t, s \right)$ is
not defined. Moreover due to the non monotonicity of $q \left( t \right)$ the
matrices $T \left( t, s \right)$ cannot always be interpreted as stochastic
matrices. Of course in the Markovian case, corresponding to
Eq.~(\ref{eq:expwtd}) and therefore to $q \left( t \right) = \exp \left( - 2
\lambda t \right)$, all these features are recovered.

The variety of possible behaviour is best clarified by considering explicit
expressions for the waiting time distribution $f \left( t \right)$ which
determines the process once the stochastic matrix $\Pi$ is given. Quite
general expressions of waiting time distribution can be obtained by
considering convex mixtures or convolutions of exponential waiting time
distributions with equal or different parameters, whose Laplace transform is
given by rational functions {\cite{Cox1965,Medhi1994}}. To better understand
the dynamics generated by the maps Eq.~(\ref{eq:tpi12}) and
Eq.~(\ref{eq:tpi1}) in the following examples note that for \ $\pi = 1 / 2$
the matrix $\Pi$ is idempotent, sending each probability vector to the uniform
distribution, while for $\pi = 1$ one has $\Pi^{2 n} = \mathbbm{1}$, and the
action of the bistochastic matrix consists in swapping the two elements of the
probability vector.

\subsubsection{Convolution of exponential waiting time distributions}

\begin{figure}[tb]
  \resizebox{80mm}{!}{\includegraphics{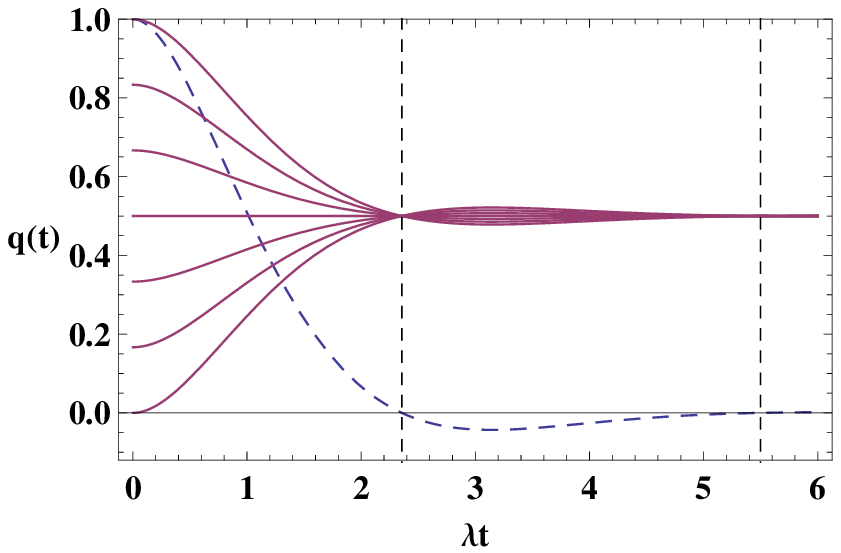}}{\hspace{1em}}\resizebox{80mm}{!}{\includegraphics{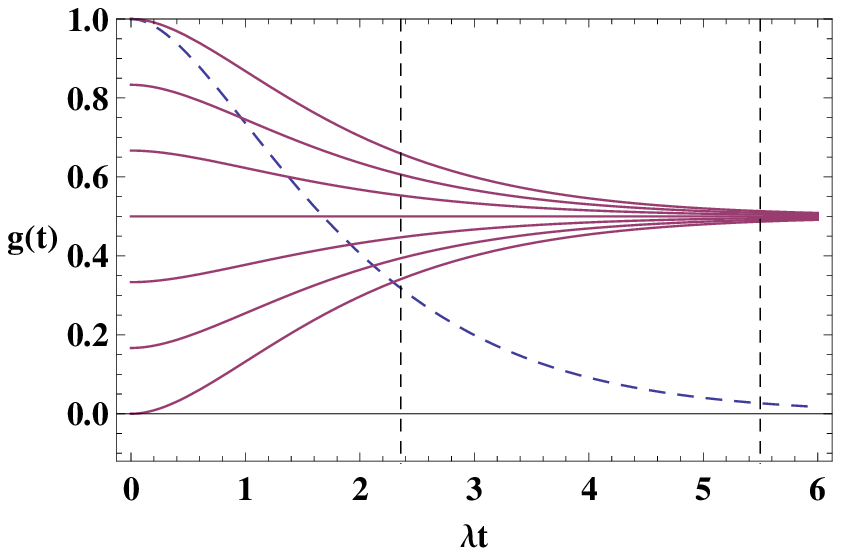}}
  \caption{\label{fig:cross1}(Color online) (top) Plot of the function $q$
  (dashed line) given by Eq.~(\ref{eq:qdiff}) as a function of $\lambda t$ for
  the convolution of two equal exponential waiting time distributions, for the
  case of a semi-Markov process with $\pi = 1$, together with a few
  trajectories (continuous lines) for the upper element $w \left( t \right)$
  of the one-point probability according to Eq.~(\ref{eq:paramet}). The
  initial data $w \left( 0 \right)$ are uniformly taken between 0 and 1. When
  $q \left( t \right)$ goes through zero the trajectories cross. At these
  points $\left| q \left( t \right) \right|$ starts growing, which indicates
  the failure of P-divisibility of the time evolution as defined in
  Eq.~(\ref{eq:cdiv}). (bottom) Plot of the function $g$ \ given by
  Eq.~(\ref{eq:sur}) as a function of $\lambda t$, for the same waiting time
  distribution, together with a few trajectories, corresponding to a
  semi-Markov process with $\pi = 1 / 2$. As it appears, despite sharing the
  same waiting time distribution the trajectories for this semi-Markov process
  never cross, and P-divisibility holds even if the process is non-Markovian.}
\end{figure}

\begin{figure}[tb]
  \resizebox{80mm}{!}{\includegraphics{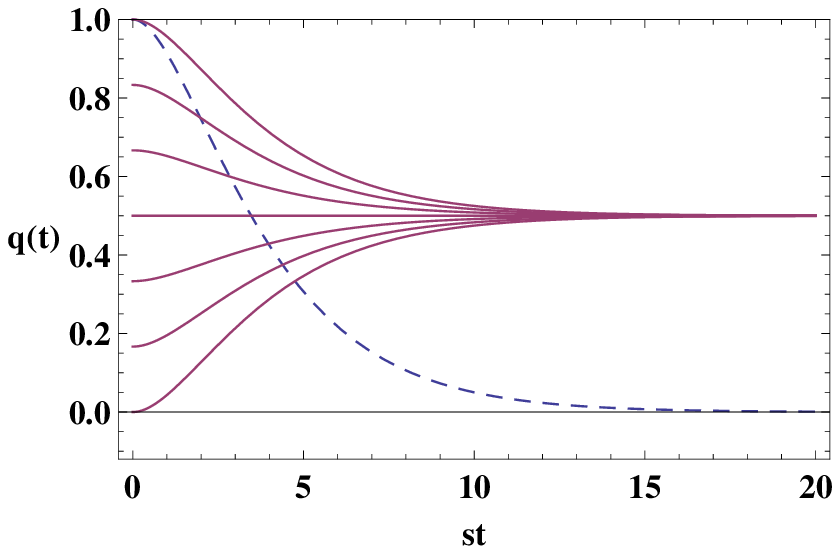}}{\hspace{1em}}\resizebox{80mm}{!}{\includegraphics{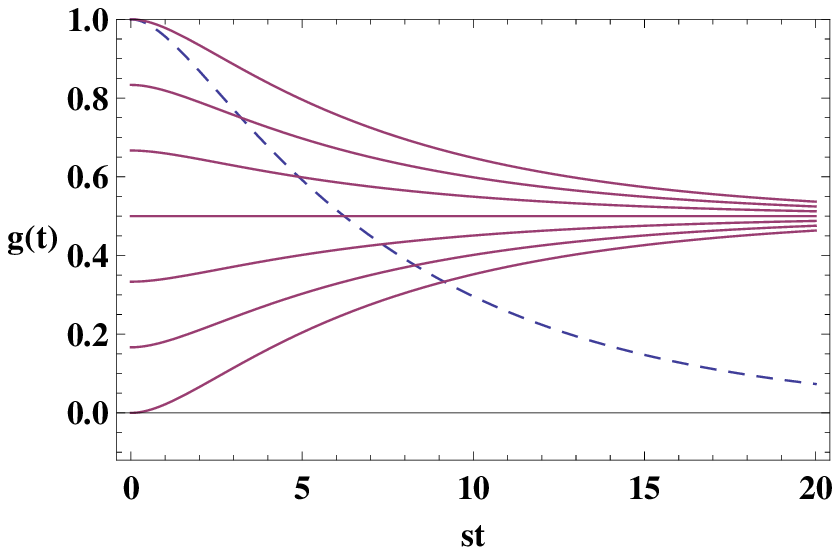}}
  \caption{\label{fig:cross2}(Color online) (top) Plot of the function $q
  \left( t \right)$ (dashed line) and of the trajectories (continuous lines)
  as in Fig.~\ref{fig:cross1}, but for the convolution of two different
  exponential waiting time distributions. We plot $q$ as a function of $st$,
  taking $p / s^2 = 0.12$, with $s$ and $p$ sum and product of the two
  parameters characterizing the exponential waiting time distributions as in
  Eq.~(\ref{eq:xi}). For this case the quantity $\chi$ given by
  Eq.~(\ref{eq:chi}) is positive, ensuring monotonicity of $q \left( t
  \right)$. (bottom) Plot of $g \left( t \right)$ and the corresponding
  trajectories for the same waiting time distribution.}
\end{figure}

The behavior of the quantity $q \left( t \right)$ can be explicitly assessed
for the case of a waiting time distribution $\mathsf{f} \left( t \right)$
given by the convolution of two exponential waiting time distributions. Let us
first consider the case in which the waiting time distributions share the same
parameter $\lambda$, so that $\mathsf{f} = f \ast f$ with $f$ as in
Eq.~(\ref{eq:expwtd}), corresponding to
\begin{eqnarray}
  \mathsf{f} \left( t \right) & = & \lambda^2 t^{} \mathe^{- \lambda t} 
  \label{eq:fsed}
\end{eqnarray}
and therefore
\begin{eqnarray}
  \mathsf{g} \left( t \right) & = & \left( 1 + \lambda t \right) \mathe^{-
  \lambda t} .  \label{eq:gsed}
\end{eqnarray}
This is a special case of the Erlang distribution {\cite{Medhi1994}}, leading
to
\begin{eqnarray}
  q \left( t \right) & = & \mathe^{- \lambda t} \left[ \cos \left( \lambda t
  \right) + \sin \left( \lambda t \right) \right],  \label{eq:qsed}
\end{eqnarray}
which oscillates and crosses zero at isolated points, so that the matrices $T
\left( t, s \right)$ are not defined at these points and cannot be always
interpreted as stochastic matrices, since their entries can become negative.
This behavior is exhibited in Fig.~\ref{fig:cross1}, where the quantity Eq.~(\ref{eq:qsed}) is plotted together with
the trajectories of the probability vector with different initial conditions.
We also consider the behavior of $\mathsf{g} \left( t \right)$ and of the
trajectories for the same waiting time distribution but a semi-Markov process
with stochastic matrix fixed by $\pi = 1 / 2$. The probability vector is of
the form
\begin{eqnarray}
  \tmmathbf{p} \left( t \right) & = & \left(\begin{array}{c}
    w \left( t \right)\\
    1 - w \left( t \right)
  \end{array}\right),  \label{eq:paramet}
\end{eqnarray}
so that its trajectories are displayed showing $w \left( t \right)$, where
\begin{eqnarray}
  w \left( t \right) & = & \frac{1}{2} \left[ 1 - q \left( t \right) + 2 q
  \left( t \right) w \left( 0 \right) \right]  \label{eq:wtq}
\end{eqnarray}
or
\begin{eqnarray}
  w \left( t \right) & = & \frac{1}{2} \left[ 1 - \mathsf{g} \left( t \right)
  + 2 \mathsf{g} \left( t \right) w \left( 0 \right) \right]  \label{eq:wtg}
\end{eqnarray}
in the two cases $\pi = 1$ and $\pi = 1 / 2$ respectively. Note how for $\pi =
1$ the different trajectories tend to group together and then separate again
depending on the behaviour of $q \left( t \right)$. A more general situation
is given by $\mathsf{f} = f_1 \ast f_2$, where each $f_i$ is of the form
Eq.~(\ref{eq:expwtd}) with parameter $\lambda_i$, so that one has
\begin{eqnarray}
  \mathsf{f} \left( t \right) & = & 2 \frac{p}{s} \mathe^{- \frac{1}{2} st}
  \frac{1}{\xi} \tmop{Sinh} \left( \frac{st}{2} \xi \right)  \label{eq:fged}
\end{eqnarray}
and correspondingly
\begin{eqnarray}
  \mathsf{g} \left( t \right) & = & \mathe^{- \frac{1}{2} st} \left[
  \tmop{Cosh} \left( \frac{st}{2} \xi \right) + \frac{1}{\xi} \tmop{Sinh}
  \left( \frac{st}{2} \xi \right) \right],  \label{eq:gged}
\end{eqnarray}
where we have set
\begin{eqnarray}
  s & = & \lambda_1 + \lambda_2 \nonumber\\
  p & = & \lambda_1 \lambda_2 \nonumber\\
  \xi & = & \sqrt{1 - 4 \frac{p}{s^2}} .  \label{eq:xi}
\end{eqnarray}
The function $q \left( t \right)$ determining the matrices $T \left( t, s
\right)$ is now given by
\begin{eqnarray}
  q \left( t \right) & = & \mathe^{- \frac{1}{2} st} \left[ \tmop{Cosh} \left(
  \frac{st}{2} \chi \right) + \frac{1}{\chi} \tmop{Sinh} \left( \frac{st}{2}
  \chi \right) \right]  \label{eq:qged}
\end{eqnarray}
with
\begin{eqnarray}
  \chi & = & \sqrt{1 - 8 \frac{p}{s^2}} .  \label{eq:chi}
\end{eqnarray}
The expression given by Eq.~(\ref{eq:qged}) shows an oscillatory behaviour if
$\chi$ becomes imaginary, for $3 - 2 \sqrt{2} \leqslant \lambda_1 / \lambda_2
\leqslant 1 / \left( 3 - 2 \sqrt{2} \right)$, while it is a positive monotonic
function of $t$ otherwise. The latter situation is considered in
Fig.~\ref{fig:cross2}.

\subsubsection{Mixture of exponential waiting time distributions}

\begin{figure}[tb]
  \resizebox{80mm}{!}{\includegraphics{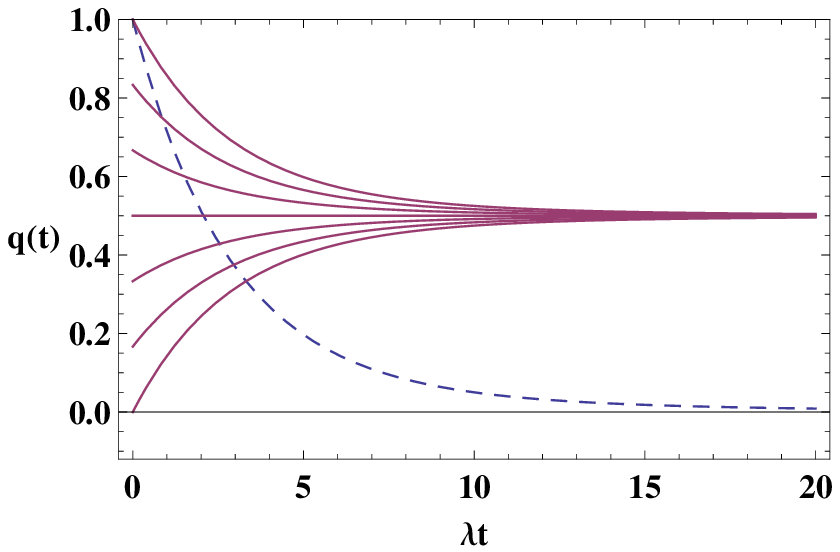}}{\hspace{1em}}\resizebox{80mm}{!}{\includegraphics{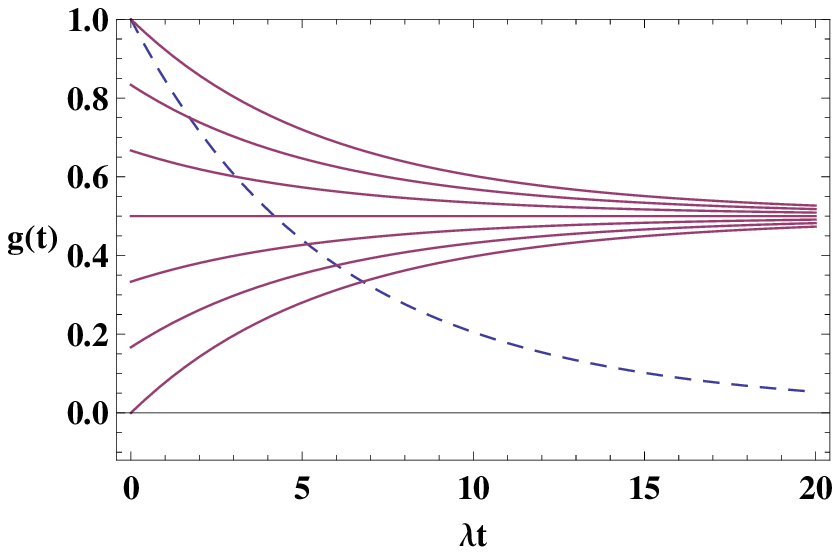}}
  \caption{\label{fig:cross3}(Color online) (top) Plot of $q \left( t
  \right)$ (dashed line) and the trajectories (continuous lines) as in
  Fig.~\ref{fig:cross1}, but for the convex mixture of two different
  exponential waiting time distributions. We plot $q$ as as a function of
  $\lambda t$, taking $\lambda_1 = a_1 \lambda$ and $\lambda_2 = a_2 \lambda$,
  with $a_1 = .1$ and $a_2 = .2$, together with mixing parameter $\mu = 0.3$.
  (bottom) Plot of $g \left( t \right)$ and the corresponding trajectories for
  the same waiting time distribution.}
\end{figure}

For the case of a convex mixture of two exponential distributions on the
contrary the trajectories never cross, and the matrices $T \left( t, s
\right)$ always are well-defined stochastic matrices. Indeed this can be seen
from
\begin{eqnarray}
  \mathsf{f} \left( t \right) & = & \mu f_1 \left( t \right) + \left( 1 - \mu
  \right) f_2 \left( t \right),  \label{eq:fcm}
\end{eqnarray}
with $0 \leqslant \mu \leqslant 1$, so that
\begin{eqnarray}
  \mathsf{g} \left( t \right) & = & \mu \mathe^{- \lambda_1 t} + \left( 1 -
  \mu \right) \mathe^{- \lambda_2 t}  \label{eq:gcm}
\end{eqnarray}
and
\begin{align}
q \left( t \right)  &=  \mathe^{- \frac{1}{2} \left( \lambda_1 + \lambda_2
  + \langle \lambda \rangle \right) t} \label{eq:qcm}
\\
& \hphantom{=}\times
\left[ \tmop{Cosh} \left(
  \frac{\bar{\lambda} t}{2} \right) + \frac{\left( \lambda_1 + \lambda_2 - 3
  \langle \lambda \rangle \right)}{\bar{\lambda}} \tmop{Sinh} \left(
  \frac{\bar{\lambda} t}{2} \right) \right],  \nonumber 
\end{align}
with
\begin{eqnarray}
  \langle \lambda \rangle & = & \mu \lambda_1 + \left( 1 - \mu \right)
  \lambda_2  \label{eq:lmean}
\end{eqnarray}
the mean rate and
\begin{eqnarray}
  \bar{\lambda} & = & \sqrt{\left( \lambda_1 + \lambda_2 + \langle \lambda
  \rangle \right)^2 - 8 \lambda_1 \lambda_2} .  \label{eq:lbar}
\end{eqnarray}
This case is considered in Fig.~\ref{fig:cross3}.

It should be noticed that in all these situations the process is
non-Markovian, but P-divisibility of the time evolution $\Lambda \left( t, 0
\right)$ in the sense of Eq.~(\ref{eq:cdiv}) still holds in some cases. For a
semi-Markov process with semi-Markov matrix as in Eq.~(\ref{eq:qbisto}) and
$\pi = 1 / 2$, P-divisibility always holds, in particular as shown in the
examples the trajectories never cross and $\left| q \left( t \right) \right|$
never grows. For $\pi = 1$ on the contrary the behaviour depends on the
waiting time distribution, which determines whether or not the quantity $q
\left( t \right)$ shows an oscillating behavior, implying that the
trajectories start getting closer till they cross and then get apart once
again.

\subsection{Kolmogorov distance}

It is interesting to relate this behavior to the time dependence of the
Kolmogorov distance among probability distributions arising from different
initial states. The Kolmogorov distance between two probability vectors can be
written as {\cite{Fuchs1999a,Nielsen2000}}
\begin{eqnarray}
  D_K \left( \tmmathbf{p}^1 \left( t \right), \tmmathbf{p}^2 \left( t \right)
  \right) & = & \frac{1}{2} \sum_i \left| p_i^1 \left( t \right) - p_i^2
  \left( t \right) \right| .  \label{eq:dk}
\end{eqnarray}
If the map $\Lambda \left( t, 0 \right)$ is P-divisible in the sense of
Eq.~(\ref{eq:cdiv}), then the Kolmogorov distance is a monotonic decreasing
function of time. Indeed by the two basic properties of a stochastic matrix,
the positivity of its entries and the fact that each row sum up to one, one
has for $t \geqslant s \geqslant 0$
\begin{eqnarray}
  D_K \left( \tmmathbf{p}^1 \left( t \right), \tmmathbf{p}^2 \left( t \right)
  \right) & = & \frac{1}{2} \sum_i \left| \sum_k \Lambda \left( t, s
  \right)_{ik} \left( \tmmathbf{p}^1 \left( s \right) -\tmmathbf{p}^2 \left( s
  \right) \right)_k \right| \nonumber\\
  & \leqslant & \frac{1}{2} \sum_i \sum_k \Lambda \left( t, s \right)_{ik}
  \left| \left( \tmmathbf{p}^1 \left( s \right) -\tmmathbf{p}^2 \left( s
  \right) \right)_k \right| \nonumber\\
  & = & \frac{1}{2} \sum_k \left| \left( \tmmathbf{p}^1 \left( s \right)
  -\tmmathbf{p}^2 \left( s \right) \right)_k \right| \nonumber\\
  & = & D_K \left( \tmmathbf{p}^1 \left( s \right), \tmmathbf{p}^2 \left( s
  \right) \right) .  \label{eq:mkd}
\end{eqnarray}
This holds true independently of the fact that the underlying classical
process is Markovian or not, it only depends on the fact the one-point
probabilities can be related at different times via stochastic matrices.

In a generic non-Markovian situation the Kolmogorov distance can both show a
monotonic decreasing behavior as well as revivals. Indeed, focusing on the
examples considered above, for a semi-Markov matrix as in
Eq.~(\ref{eq:qbisto}) and $\pi = 1 / 2$ one has
\begin{eqnarray}
  D_K \left( \tmmathbf{p}^1 \left( t \right), \tmmathbf{p}^2 \left( t \right)
  \right) & = & g \left( t \right) D_K \left( \tmmathbf{p}^1 \left( 0 \right),
  \tmmathbf{p}^2 \left( 0 \right) \right),  \label{eq:dkg}
\end{eqnarray}
while for $\pi = 1$ one has
\begin{eqnarray}
  D_K \left( \tmmathbf{p}^1 \left( t \right), \tmmathbf{p}^2 \left( t \right)
  \right) & = & \left| q \left( t \right) \right| D_K \left( \tmmathbf{p}^1
  \left( 0 \right), \tmmathbf{p}^2 \left( 0 \right) \right) .  \label{eq:dkq}
\end{eqnarray}
Thus, while for $\pi = 1 / 2$ the Kolmogorov distance is a monotonic
contraction for any waiting time distribution, thanks to the fact that $g
\left( t \right)$ is a survival probability, for $\pi = 1$ the distance among
distributions can show revivals depending on the explicit expression of $q
\left( t \right)$, as can be seen from Fig.~\ref{fig:cross1} for the case of
the convolution of two exponential waiting time distributions with the same
parameter.

We have thus studied, by means of explicit examples, the behavior of the
probability vector or one-point probability $\tmmathbf{p} \left( t \right)$ of
a classical process. In particular, we have seen that while for a Markovian
process P-divisibility is always granted and the Kolmogorov distance is a
monotone contraction, non-Markovianity can spoil these features, even though
neither the lack of P-divisibility nor the growth of the Kolmogorov distance
can be taken as necessary signatures of non-Markovianity. This is not
surprising due to the fact that the non-Markovianity of a classical process
cannot be traced back to the behavior of the one-point probabilities, since it
involves all $n$-point probabilities.

\section{Quantum non-Markovian processes\label{sec:qnm}}

We now come back to the quantum realm, studying a class of quantum dynamics
which have a clearcut physical meaning, allowing both for the evaluation and
the comparison of two recently introduced measures of non-Markovianity for the
quantum case, and for a direct connection with the classical situation
analyzed in Sect.~\ref{sec:cnm}. Indeed, while in the classical case one has a
well settled definition of non-Markovianity for a stochastic process, which
can be used to speak of Markovianity or non-Markovianity of the time evolution
of the associated probability distributions, equivalent characterizations for
the quantum case have been proposed only recently. As follows from the
discussion in Sect.~\ref{sec:cnm}, such approaches cope by necessity with the
behavior of the one-point probabilities only, which can be obtained from the
statistical operator $\rho \left( t \right)$, since a definition involving the
whole hierarchy of $n$-point probabilities cannot be introduced without
explicit reference to a particular choice of measurement scheme. Note that in
the study of quantum dynamics one speaks about measures of non-Markovianity,
since apart from clarifying what is the signature of non-Markovianity, so as
to define it and therefore make it detectable, one would like to quantify the
degree of non-Markovianity of a given dynamics.

The two measures that we will consider here {\cite{Breuer2009b,Rivas2010a}}
actually respectively rely on the violation of the quantum analog of the
classical properties of P-divisibility and monotonic decrease in time of the
Kolmogorov distance, that we have considered in Sect.~\ref{sec:cnm} and hold
true for the Markovian case, while also other approaches have been considered
{\cite{Wolf2008a}}. Note that the violation of these properties in the
classical case provide a sufficient but not necessary condition to detect a
non-Markovian process, as clarified in the examples considered in
Sect.~\ref{sec:csmp}.

\subsection{Quantum semi-Markov processes\label{sec:qsmp}}

As in the classical case, in order to study the Markovian or non-Markovian
features of a quantum dynamics we consider a class of time evolutions which
allow for an explicit treatment and the connection to a classical counterpart.
Let us study dynamics given by master equations of the form
{\cite{Budini2004a}}
\begin{eqnarray}
  \frac{\mathd}{\tmop{dt}} \rho \left( t \right) & = & \int^t_0 d \tau k
  \left( t - \tau \right) \left[ \mathcal{E} - \mathbbm{1} \right] \rho \left(
  \tau \right),  \label{eq:ms}
\end{eqnarray}
where $\mathcal{E}$ is a CPT map and $k \left( t \right)$ a memory kernel
associated to a waiting time distribution $f \left( t \right)$ according to
Eq.~(\ref{eq:k}). Such master equations generate by construction completely
positive dynamics, which provide a quantum counterpart of classical
semi-Markov processes {\cite{Breuer2008a,Breuer2009a}}. Indeed, as it can be
checked, the solution of Eq.~(\ref{eq:ms}) can be written as
\begin{eqnarray}
  \rho \left( t \right) & = & \Phi \left( t, 0 \right) \rho \left( 0 \right)
  \nonumber\\
  & = & \sum^{\infty}_{n = 0} p_n (t) \mathcal{E}^n \rho \left( 0 \right), 
  \label{eq:gspl}
\end{eqnarray}
with $p_n (t)$ fixed by $f \left( t \right)$ according to Eq.~(\ref{eq:pn}),
while $\mathcal{E}^n$ denotes the $n$-fold composition of the map
$\mathcal{E}$. The map $\Phi \left( t, 0 \right)$ is thus completely positive
by construction and can be interpreted as the repeated action of the CPT map
$\mathcal{E}$, implementing a quantum operation, distributed in time according
to the waiting time distribution $f \left( t \right)$. For the case of an
exponential waiting time distribution the memory kernel is a delta function
and one goes back to a semigroup dynamics. In Sect.~\ref{sec:cnm} we
considered semi-Markov processes with a semi-Markov matrix of the form
Eq.~(\ref{eq:qbisto}), with arbitrary $f \left( t \right)$ and $\Pi$ a
bistochastic matrix, and where Markovianity or non-Markovianity of the process
depended only on the choice of $f \left( t \right)$, while P-divisibility and
behavior of the Kolmogorov distance did depend on both $f \left( t \right)$
and $\Pi$. In the quantum setting we also leave $f \left( t \right)$ arbitrary
and consider bistochastic CPT maps, in the sense that $\mathcal{E \left[
\mathbbm{1} \right] =} \mathbbm{1}$, so that also $\Phi \left( t, 0 \right)$
is bistochastic, preserving both the trace and the identity.

\subsection{Dephasing dynamics and time-local equation}

A purely quantum dynamics, only affecting coherences, is obtained considering
the CPT map
\begin{eqnarray}
  \mathcal{E}^{}_z \rho & = & \sigma_z \rho \sigma_z,  \label{eq:ez}
\end{eqnarray}
which satisfies $\mathcal{E}_z^{2 n} = \mathbbm{1}$ and $\mathcal{E}_z^{2 n +
1} = \mathcal{E}_z$, so that one has{\hspace*{\fill}}
\begin{eqnarray}
  \rho \left( t \right) & = & p_{\tmop{even}} \left( t \right) \rho \left( 0
  \right) - p_{\tmop{odd}} \left( t \right) \sigma_z \rho \left( 0 \right)
  \sigma_z  \label{eq:mapez}\\
  & = & \left( \begin{array}{cc}
    \rho_{11} (0) & q (t) \rho_{10} (0)\\
    q (t) \rho_{01} (0) & \rho_{00} (0)
  \end{array} \right), \nonumber
\end{eqnarray}
recalling the definition Eq.~(\ref{eq:qdiff}) of $q \left( t \right)$ and
considering matrix elements in the basis of eigenvectors of $\sigma_z$. Before
addressing the issue of characterization of these dynamics, it is of interest
to recast the integrodifferential master equation Eq.~(\ref{eq:ms}) in a
time-convolutionless form. Indeed, while Markovianity or non-Markovianity is a
property of the solution $\rho \left( t \right)$, rather then of the equation,
it is quite important to read the signatures of a non-Markovian behavior from
the equations themselves, and this task turns out to be much easier when the
equations are written in time-local form. To rewrite the master equation
Eq.~(\ref{eq:ms}) in time-local form we follow the constructive approach used
in {\cite{Smirne2010b}}. We thus obtain the time-convolutionless generator,
which is formally given by $\dot{\Phi} \left( t, 0 \right) \Phi^{- 1} \left(
t, 0 \right)$, by a representation of the map as a matrix with respect to a
suitable operator basis in $\mathbbm{C}^2$, given by $\left\{ X_i \right\} =
\left\{ \frac{1}{\sqrt{2}} \mathbbm{1}, \frac{1}{\sqrt{2}} \sigma_i
\right\}_{i = 1, 2, 3}$, orthonormal according to $\tmop{Tr}_S \left\{ X_i X_j
\right\} = \delta_{ij}$. The matrix $F = \left( F_{kl} \right)$ associated to
the map $\Phi$ is then determined as
\begin{eqnarray}
  F_{kl} & = & \tmop{Tr}_S \left\{ X^{\dag}_k \Phi \left[ X_l \right]
  \right\}, \nonumber
\end{eqnarray}
so that
\begin{eqnarray}
  \Phi \left[ \rho \right] & = & \sum_{kl} F_{kl} \tmop{Tr}_S \left\{
  X_l^{\dag} \rho \right\} X_k . \nonumber
\end{eqnarray}
The matrix $F$ for our map is given by
\begin{eqnarray}
  F \left( t, 0 \right) & = & \tmop{diag} \left( 1, q \left( t \right), q
  \left( t \right), 1 \right)  \label{eq:ft1}
\end{eqnarray}
and accordingly the time-convolutionless master equation simply reads
\begin{eqnarray}
  \frac{\mathd}{\tmop{dt}} \rho \left( t \right) & = & \gamma \left( t \right)
  \mathcal{L}_z \left[ \rho \left( t \right) \right],  \label{eq:tclz}
\end{eqnarray}
where we have a single quantum channel given by
\begin{eqnarray}
  \mathcal{L}_z \left[ \rho \right] & = & \sigma_z \rho \sigma_z - \rho 
  \label{eq:lz}
\end{eqnarray}
and the time dependent rate $\gamma \left( t \right)$ is
\begin{eqnarray}
  \gamma \left( t \right) & = & - \frac{1}{2} \frac{\dot{q} \left( t
  \right)}{q \left( t \right)}  \label{eq:gammat}\\
  & = & - \frac{1}{2} \frac{\mathd}{\tmop{dt}} \log \left| q \left( t \right)
  \right| . \nonumber
\end{eqnarray}

\subsubsection{Divisibility}

Relying on the matrix representation of the map $\Phi \left( t, 0 \right)$
given by Eq.~(\ref{eq:ft1}) we are now in the position to study its
divisibility according to Eq.~(\ref{eq:qdiv}). In particular the map $\Phi
\left( t, 0 \right)$ will turn out to be P-divisible if the matrices
\begin{eqnarray}
  F \left( t, s \right) & = & \tmop{diag} \left( 1, \frac{q \left( t
  \right)}{q \left( s \right)}, \frac{q \left( t \right)}{q \left( s \right)},
  1 \right)  \label{eq:fts1}
\end{eqnarray}
obtained as in Eq.~(\ref{eq:trick}) represent a positive map $\Phi \left( t, s
\right)$ for any $t \geqslant s \geqslant 0$, which is the case provided
\begin{eqnarray}
  \left| \frac{q \left( t \right)}{q \left( s \right)} \right| & \leqslant &
  1.  \label{eq:p1}
\end{eqnarray}
This condition is satisfied if $\left| q \left( t \right) \right|$ is a
monotonic decreasing function, and therefore the time dependent rate $\gamma
\left( t \right)$ is always positive. In order to assess when CP-divisibility
holds, one can consider positivity of the associated Choi matrix
{\cite{Choi1975a}}, which still leads to the constraint (\ref{eq:p1}). It
follows that for this model CP-divisibility and P-divisibility are violated at
the same time, whenever $\left| q \left( t \right) \right|$ increases, so that
$\gamma \left( t \right)$ becomes negative. Thus, as discussed in
{\cite{Laine2010a}}, for the case of a single quantum channel positivity of
the time dependent rate ensures CP-divisibility of the time evolution, which
is violated if $\gamma \left( t \right)$ becomes negative at some point.

\subsubsection{Measures of non-Markovianity}

We can now evaluate the measures of non-Markovianity for this model according
to both approaches devised in {\cite{Breuer2009b}} and in {\cite{Rivas2010a}}.

The first approach by Breuer, Laine and Piilo relies on the study of the
behavior in time of the trace distance among different initial states. The
trace distance among quantum states quantifies their distinguishability
{\cite{Fuchs1999a,Nielsen2000}}, and is defined in terms of their distance
with respect to the trace class norm, thus providing the natural quantum
analog of the Kolmogorov distance
\begin{eqnarray}
  D \left( \rho^1 \left( t \right), \rho^2 \left( t \right) \right) & = &
  \frac{1}{2} \| \rho^1 \left( t \right) - \rho^2 \left( t \right) \|_1 . 
  \label{eq:td}
\end{eqnarray}
For our case it reads
\begin{eqnarray}
  D \left( \rho^1 \left( t \right), \rho^2 \left( t \right) \right) & = &
  \sqrt{\Delta_p \left( 0 \right)^2 + \left| \Delta_c \left( 0 \right)
  \right|^2 q^2 \left( t \right)},  \label{eq:td1}
\end{eqnarray}
where we have set
\begin{eqnarray}
  \Delta_p \left( 0 \right) & = & \rho^1_{11} \left( 0 \right) - \rho^2_{00}
  \left( 0 \right)  \label{eq:pop}\\
  \Delta_c \left( 0 \right) & = & \rho^1_{10} \left( 0 \right) - \rho^2_{10}
  \left( 0 \right)  \label{eq:coh}
\end{eqnarray}
for the differences of the populations and the coherences between two given
initial states $\rho^1 \left( 0 \right)$ and $\rho^2 \left( 0 \right)$
respectively. Its time derivative is then given by
\begin{equation}
  \sigma \left( t, \rho^{1, 2} \left( 0 \right) \right)  =  \frac{\left|
  \Delta_c \left( 0 \right) \right|^2}{\sqrt{\Delta_p \left( 0 \right)^2 +
  \left| \Delta_c \left( 0 \right) \right|^2 q^2 \left( t \right)}} q \left( t
  \right) \dot{q} \left( t \right),  \label{eq:s1}
\end{equation}
so that the trace distance among states can indeed grow provided $q \left( t
\right)$ and $\dot{q} \left( t \right)$ have the same sign, so that $\left| q
\left( t \right) \right|$ does increase. Thus the map has a positive measure
of non-Markovianity whenever P-divisibility or equivalently CP-divisibility is
broken.

In order to obtain the measure of non-Markovianity for this time evolution
according to {\cite{Breuer2009b}} one has to integrate the quantity $\sigma
\left( t, \rho^{1, 2} \left( 0 \right) \right)$ over the time region, let us
call it $\Omega_+$, where it is positive, and then maximize the result over
all possible initial pairs of states. The region $\Omega_+$ now corresponds to
the time intervals where $\left| q \left( t \right) \right|$ increases, and
the maximum is obtained for initial states such that $\Delta_p \left( 0
\right) = 0$ and $\Delta_c \left( 0 \right) = 1$, so that we have the
following explicit expression for the measure of non-Markovianity
\begin{eqnarray}
  \mathcal{N} \left( \Phi \right) & = & \int_{\Omega_+} \mathd t
  \frac{\mathd}{\mathd t}  \left| q \left( t \right) \right| 
  \label{eq:mblp}\\
  & = & \sum_i \left( \left| q \left( b_i \right) \right| - \left| q \left(
  a_i \right) \right| \right), \nonumber
\end{eqnarray}
where we have set $\Omega_+ = \bigcup_i \left( a_i, b_i \right)$. A couple
of states which maximize the growth of the trace distance is given in this
case by the pure states $\rho^{1, 2} \left( 0 \right) = | \psi_{\pm} \rangle
\langle \psi_{\pm} |$, with $\left. | \psi_{\pm} \rangle = \left( 1 / \sqrt{2}
\right) \left( |0 \right\rangle \pm |1 \rangle \right)$.

The approach by Rivas, Huelga and Plenio instead focuses on the
CP-divisibility of the map. While for this model this requirement for
non-Markovianity is satisfied at the same time as the growth of the trace
distance, the effect is quantified in a different way. Indeed these authors
quantify non-Markovianity as the integral
\begin{eqnarray}
  \mathcal{I} \left( \Phi \right) & = & \int_{\mathbbm{R_+}} \mathd
  t\mathfrak{g} \left( t \right)  \label{eq:defrhp1}
\end{eqnarray}
where the quantity $\mathfrak{g} \left( t \right)$ is given by
\begin{eqnarray}
  \mathfrak{g} \left( t \right) & = & \lim_{\epsilon \rightarrow 0}
  \frac{\frac{1}{2} \| \Phi_{\tmop{Choi}} \left( t, t + \epsilon \right) \|_1
  - 1}{\epsilon},  \label{eq:defrhp2}
\end{eqnarray}
where $\Phi_{\tmop{Choi}}$ denotes the Choi matrix associated to the map
$\Phi$, and it is different from zero only when CP-divisibility is broken. For
the case at hand one has
\begin{eqnarray}
  \mathcal{I} \left( \Phi \right) & = & \int_{\Omega_+} \mathd t
  \frac{\mathd}{\tmop{dt}} \log \left| q \left( t \right) \right| 
  \label{eq:mrhp}\\
  & = & \sum_i \left( \log \left| q \left( b_i \right) \right| - \log \left|
  q \left( a_i \right) \right| \right) \nonumber\\
  & = & - 2 \int_{\Omega_+} \mathd t \gamma \left( t \right) . \nonumber
\end{eqnarray}

For this model the growth of $\left| q \left( t \right) \right|$ determines
both the breaking of CP-divisibility as well as the growth of the trace
distance, so that both approaches detect non-Markovianity at the same time,
even if they quantify it in different ways. This is however not generally
true, as observed already in {\cite{Laine2010a}} and considered in
{\cite{Haikka2011a,Chruscinski-xxx3}}. We will point to examples for the
different performance of the two measures later on. Exploiting the results of
Sect.~\ref{sec:csmp} it is now interesting to consider explicit choices of
waiting time distributions, so as to clarify the different possible behaviors.

\begin{flushleft}
  \subsubsection{Explicit example}\label{sec:ee2}
\end{flushleft}

For the case of a memoryless waiting time distribution of the form
Eq.~(\ref{eq:expwtd}), so that $k \left( t \right)$ is actually a delta
function and $q \left( t \right) = \exp \left( - 2 \lambda t \right)$,
according to Eq.~(\ref{eq:gammat}) the function $\gamma \left( t \right)$ is
simply given by the positive constant $\lambda$. Each non-Markovianity measure
is easily assessed to be zero.

\begin{figure}[tb]
  \resizebox{80mm}{!}{\includegraphics{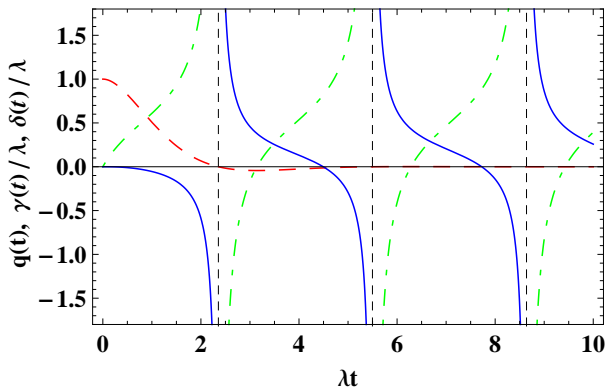}}
  \caption{\label{fig:rates1}(Color online) Plot of $q$ (red dashed line),
  $\gamma$ (green dot-dashed line) and $\delta$ (blue continuous line) defined in
  Eq.~(\ref{eq:qdiff}), Eq.~(\ref{eq:gammat}) and Eq.~(\ref{eq:defd})
  respectively, as functions of $\lambda t$ for the convolution of two equal
  exponential waiting time distributions. The vertical asymptotes denote the
  points where $q \left( t \right)$ goes through zero, so that $\gamma \left(
  t \right)$ diverges. The functions $\gamma \left( t \right)$ and $\delta
  \left( t \right)$ appear as time dependent coefficients in front of the
  various quantum channels in the time local quantum master equations given by
  Eq.~(\ref{eq:tclz}) and Eq.~(\ref{eq:tclpm}), so that their sign determines
  P-divisibility and CP-divisibility of the corresponding quantum semi-Markov
  processes, as discussed in Sect.~\ref{sec:qsmp}.}
\end{figure}

To consider non-trivial situations, non-Markovian in the classical case, let
us first assume a waiting time distribution of the form Eq.~(\ref{eq:fsed}),
arising by convolving two exponential memoryless distributions with the same
parameter. The function $q \left( t \right)$ is then given by
Eq.~(\ref{eq:qsed}), so that $\gamma \left( t \right)$ reads
\begin{eqnarray}
  \gamma \left( t \right) & = & \lambda \frac{1}{1 + \tmop{cotg} \left(
  \lambda t \right)},  \label{eq:gtz}
\end{eqnarray}
which indeed takes on both positive and negative values, diverging for
$\lambda t = \left( 3 / 4 \right) \pi$ mod $\pi$. Both functions are plotted
in Fig.~\ref{fig:rates1}. In this case the region $\Omega_+$ can be exactly determined and
is given by
\begin{eqnarray}
  \Omega_+ & = & \bigcup_{n \in \mathbbm{N}} \left( \frac{1}{\lambda} \left(
  \pi + n \pi \right), \frac{1}{\lambda} \left( \frac{3}{4} \pi + n \pi
  \right) \right) . \label{omegaplus}
\end{eqnarray}
As already observed both measures become nonzero when $\left| q \left( t
\right) \right|$ grows. The measure proposed by Breuer, Laine and Piilo
according to Eq.~(\ref{eq:mblp}) can now be exactly calculated and is given by
\begin{eqnarray}
  \mathcal{N} \left( \Phi \right) & = & \sum^{\infty}_{n = 0} \left( -
  \right)^{n + 1} \left[ q \left( \frac{\pi + n \pi}{\lambda} \right) - q
  \left( \frac{3 \pi / 4 + n \pi}{\lambda} \right) \right] \nonumber \label{eq:exact}\\
  & = & \frac{1}{e^{\pi} - 1},
\end{eqnarray}
which is finite and independent on $\lambda$. It is to be stressed that
considering the convolution of a higher number of exponential waiting time
distributions one obtains a higher value for this measure, according to the
fact that the overall waiting time distribution departs more and more from the
memoryless exponential case {\cite{Cox1965}}. The measure proposed by Rivas,
Huelga and Plenio instead is equal to infinity $\mathcal{I} \left( \Phi
\right) \eqcirc \infty$, due to the fact that $q \left( t \right)$ goes
through zero and therefore $\gamma \left( t \right)$ diverges. Actually
$\mathcal{I} \left( \Phi \right)$ is equal to infinity whenever the inverse
time evolution mapping fails to exist. It therefore quantifies in the same way
non-Markovianity for quite different situations, e.g. in this case waiting
time distributions given by the convolution of a different number of
exponentials.

\begin{figure}[tb]
  \resizebox{80mm}{!}{\includegraphics{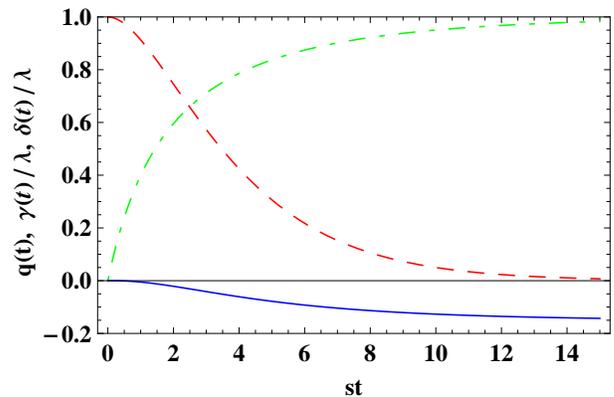}}
  \caption{\label{fig:rates2}(Color online) The same as Fig.~\ref{fig:rates1},
  but for the convolution of two different exponential waiting time
  distributions. We plot the quantities as a function of $st$, taking $p / s^2
  = 0.12$. Note that in this case $\gamma \left( t \right)$ (green dot-dashed line)
  is always positive, while $\delta \left( t \right)$ (blue continuous line) is
  always negative. The function $q \left( t \right)$ (red dashed line)
  monotonically decreases reaching the value zero only at infinity.}
\end{figure}

As a further example we consider a convolution of two different exponential
distributions, corresponding to Eq.~(\ref{eq:fged}) and Eq.~(\ref{eq:qged}),
so that one now has
\begin{eqnarray}
  \gamma \left( t \right) & = & 2 \frac{p}{s}  \frac{1}{1 + \chi \tmop{Coth}
  \left( \frac{st}{2} \chi \right)} .  \label{eq:gtz2}
\end{eqnarray}
Recalling Eq.~(\ref{eq:xi}) and Eq.~(\ref{eq:chi}) the argument of the
hyperbolic cotangent is real, so that $\gamma \left( t \right)$ always stays
positive, if $p \leqslant s^2 / 8$. In this case, despite the underlying
non-Markovian classical process, both measures of non-Markovianity are equal
to zero. The behavior of $q \left( t \right)$ and $\gamma \left( t \right)$
for this case is depicted in Fig.~\ref{fig:rates2}. When $p > s^2 / 8$, which includes the case $\lambda_1 =
\lambda_2$, $q \left( t \right)$ again oscillates between positive and
negative values, so that one has a similar behavior as before, with
$\mathcal{N} \left( \Phi \right)$ assuming a finite value and $\mathcal{I}
\left( \Phi \right) \eqcirc \infty$.

\begin{figure}[tb]
  \resizebox{80mm}{!}{\includegraphics{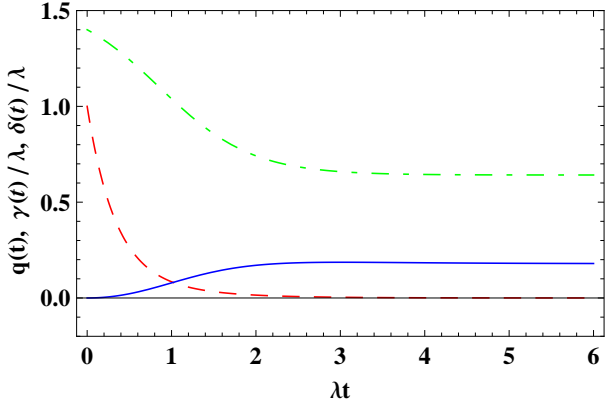}}
   \caption{\label{fig:rates3}(Color online) The same as Fig.~\ref{fig:rates1},
  but for a convex mixture of two different exponential waiting time
  distributions. We plot $q$ (red dashed line), $\gamma$ (green dot-dashed line) and
  $\delta$ (blue continuous line) as a function of $\lambda t$, taking
  $\lambda_1 = a_1 \lambda$ and $\lambda_2 = a_2 \lambda$, with $a_1 = 1$ and
  $a_2 = 6$, together with mixing parameter $\mu = 0.6$. For this kind of
  waiting time distribution all functions always stay positive, quickly
  reaching an asymptotic constant value.}
\end{figure}

Finally let us consider a convex mixture of two memoryless distributions as
given by Eq.~(\ref{eq:fcm}), so that $q \left( t \right)$ is now given by
Eq.~(\ref{eq:qcm}) and one has
\begin{equation}
  \gamma \left( t \right)  =  \langle \lambda \rangle \frac{1 + \frac{1}{4}
  \frac{\left( \lambda_1 + \lambda_2 - 3 \langle \lambda \rangle \right)
  \left( \lambda_1 + \lambda_2 + \langle \lambda \rangle \right) -
  \overline{\lambda}^2}{ \overline{\lambda} \langle \lambda \rangle}
  \tmop{Coth} \left( \frac{\bar{\lambda} t}{2} \right)}{1 + \frac{\left(
  \lambda_1 + \lambda_2 - 3 \langle \lambda \rangle \right)}{
  \overline{\lambda}} \tmop{Coth} \left( \frac{\bar{\lambda} t}{2} \right)}, 
  \label{eq:gtz3}
\end{equation}
which according to the definitions of $\langle \lambda \rangle$ and
$\overline{\lambda}$ given in Eq.~(\ref{eq:lmean}) and Eq.~(\ref{eq:lbar}) can
be checked to always take on positive values. Its behavior is given in
Fig.~\ref{fig:rates3}. In this situation both measures are equal to zero.

\subsection{Dephasing dynamics via projection}

A quantum dynamics corresponding to pure dephasing is also obtained
considering a CPT map $\mathcal{E}$ which is also a projection, that is
\begin{eqnarray}
  \mathcal{E}_{_P} \rho & = & \sigma_+ \sigma_- \rho \sigma_+ \sigma_- +
  \sigma_- \sigma_+ \rho \sigma_- \sigma_+  \label{eq:ep}
\end{eqnarray}
so that one has idempotency $\mathcal{E}^2_{_P} = \mathcal{E}_{_P}$. For this
case the analysis closely follows the one performed for $\mathcal{E}_{z}$, but
the crucial quantity instead of $q \left( t \right)$ is the survival
probability $g \left( t \right)$, similarly to the classical case with $\pi =
1 / 2$ dealt with in Sect.~\ref{sec:csmp}. The matrix $F$ is given by
\begin{eqnarray}
  F \left( t, 0 \right) & = & \tmop{diag} \left( 1, g \left( t \right), g
  \left( t \right), 1 \right)  \label{eq:ftp}
\end{eqnarray}
and the time-local master equation reads
\begin{eqnarray}
  \frac{\mathd}{\tmop{dt}} \rho \left( t \right) & = & h \left( t \right)
  \left( \mathcal{L}_{+ -} \left[ \rho \left( t \right) \right] +
  \mathcal{L}_{- +} \left[ \rho \left( t \right) \right] \right), 
  \label{eq:tclp}
\end{eqnarray}
with Lindblad operators
\begin{eqnarray}
  \mathcal{L}_{+ -} \left[ \rho \right] & = & \sigma_+ \sigma_- \rho \sigma_+
  \sigma_- - \frac{1}{2} \left\{ \sigma_+ \sigma_-, \rho \right\} 
  \label{eq:lpm}
\end{eqnarray}
and similarly for $\mathcal{L}_{- +}$. The quantity $h \left( t \right)$ is
given by
\begin{eqnarray}
  h \left( t \right) & = & \frac{f \left( t \right)}{g \left( t \right)} 
  \label{eq:h}\\
  & = & - \frac{\mathd}{\tmop{dt}} \log g \left( t \right), \nonumber
\end{eqnarray}
which provides the so called hazard rate function associated to the waiting
time distribution $f \left( t \right)$, given by the ratio of waiting time
distribution and survival probability. It gives information on the probability
for the first jump to occur right after time $t$ {\cite{Ross2007}}. Note that
the survival probability is a positive monotonously decreasing function, and
the hazard rate function is always positive. As a result CP-divisibility
always holds, so that both non-Markovianity measures are equal to zero,
whatever the waiting time distribution is.

\subsection{Dissipative dynamics and time-local equation}

The choice of CPT map considered above, corresponding to pure dephasing, shows
how different probability densities for the waiting time, corresponding to
different distributions of the action of the quantum operation in time, can
lead to dynamics whose measures of non-Markovianity can be both positive or
zero, irrespective of the fact that the only memoryless waiting time
distribution is the exponential one. In this case, however, there is no direct
connection to a classical dynamics, since only the coherences evolve in time.
Another natural choice of CPT map which leads to a non trivial dynamics for
the populations is given by
\begin{eqnarray}
  \mathcal{E}_{\pm} \rho & = & \sigma_- \rho \sigma_+ + \sigma_+ \rho
  \sigma_-,  \label{eq:epm}
\end{eqnarray}
which satisfies $\mathcal{E}_{\pm}^{2 n} = \mathcal{E}^2_{\pm}$ and
$\mathcal{E}_{\pm}^{2 n + 1} = \mathcal{E}_{\pm}$, so that one can obtain the
explicit representation
\begin{eqnarray}
   \label{eq:1}
   \rho_{11} \left( t \right) &=& p_{\tmop{{even}}} \left( t \right) \rho_{11} (0) + p_{\tmop{{odd}}} \left( t
    \right) \rho_{00} \left( 0 \right) \nonumber \\
\rho_{10} \left( t \right) &=& g (t) \rho_{10} (0)\nonumber \\
\rho_{01} \left( t \right) &=& g (t) \rho_{01} (0)\nonumber \\
\rho_{00} \left( t \right) &=& p_{\tmop{{odd}}} \left( t \right) \rho_{11} \left( 0
    \right) + p_{\tmop{{even}}} \left( t \right) \rho_{00} (0)
\end{eqnarray}
where $g (t)$ denotes as usual the survival probability. As in the previous
case we can obtain the matrix $F$ representing the action of the map with
respect to the chosen basis of operators in $\mathbbm{C}^2$, now given by
\begin{eqnarray}
  F \left( t, 0 \right) & = & \tmop{diag} \left( 1, g \left( t \right), g
  \left( t \right), q \left( t \right) \right)  \label{eq:ft2}
\end{eqnarray}
and accordingly the time-convolutionless master equation reads
\begin{equation}
  \frac{\mathd}{\tmop{dt}} \rho \left( t \right)  =  \gamma \left( t \right)
  \left( \mathcal{L}_+ \left[ \rho \left( t \right) \right] + \mathcal{L}_-
  \left[ \rho \left( t \right) \right] \right) + \delta \left( t \right)
  \mathcal{L}_z \left[ \rho \left( t \right) \right],  \label{eq:tclpm}
\end{equation}
where besides $\mathcal{L}_z$ as given by Eq.~(\ref{eq:lz}) the quantum
channels
\begin{equation}
  \mathcal{L}_+ \left[ \rho \right]  =  \sigma_+ \rho \sigma_- - \frac{1}{2}
  \left\{ \sigma_- \sigma_+, \rho \right\}  \label{eq:lp}
\end{equation}
and
\begin{eqnarray}
  \mathcal{L}_- \left[ \rho \right] & = & \sigma_- \rho \sigma_+ - \frac{1}{2}
  \left\{ \sigma_+ \sigma_-, \rho \right\}  \label{eq:lm}
\end{eqnarray}
appear. The time dependent rate $\gamma \left( t \right)$ is still given by
Eq.~(\ref{eq:gammat}), while the function $\delta \left( t \right)$ is given
by the difference
\begin{eqnarray}
  \delta \left( t \right) & = & \frac{1}{2} \left( h \left( t \right) - \gamma
  \left( t \right) \right),  \label{eq:defd}
\end{eqnarray}
where $h \left( t \right)$ is the so called hazard rate function introduced in
Eq.~(\ref{eq:h}), which is always positive.

\subsubsection{Divisibility}

Also in this case we can consider the divisibility properties of the time
evolution. According to the matrix representation of the map,
Eq.~(\ref{eq:trick}) now leads to
\begin{eqnarray}
  F \left( t, s \right) & = & \tmop{diag} \left( 1, \frac{g \left( t
  \right)}{g \left( s \right)}, \frac{g \left( t \right)}{g \left( s \right)},
  \frac{q \left( t \right)}{q \left( s \right)} \right),  \label{eq:fts2}
\end{eqnarray}
so that thanks to the property of the survival probability the only condition
for P-divisibility is still given by Eq.~(\ref{eq:p1}). Therefore the map is \
P-divisible whenever $\left| q \left( t \right) \right|$ is a monotonic
decreasing function. Note that this condition implies positivity of $\gamma
\left( t \right)$, and therefore of the time dependent rate in front of the
$\mathcal{L}_+$ and $\mathcal{L}_-$ channels, which affect the dynamics of the
populations. In order to study CP-divisibility one has to consider the
associated Choi matrix, whose positivity is granted upon the further condition
\begin{eqnarray}
  \frac{g \left( t \right)}{g \left( s \right)} & \leqslant & \frac{1}{2}
  \left( 1 + \frac{q \left( t \right)}{q \left( s \right)} \right), 
  \label{eq:cp2}
\end{eqnarray}
which sets a non trivial requirement, implying positivity of the function
$\delta \left( t \right)$ which provides the coefficient of the purely quantum
channel $\mathcal{L}_z$. Thus CP-divisibility is violated if and only if at
least one of the prefactors in the time-local form Eq.~(\ref{eq:tclpm})
becomes negative. Note however that in this case, due to the presence of
different quantum channels, P-divisibility and CP-divisibility are not
necessarily violated together, since it can well happen that $\gamma \left( t
\right)$ stays positive, but $\delta \left( t \right)$ takes on negative
values. As discussed in the examples below and shown in Fig.~\ref{fig:rates2},
for a suitable choice of waiting time distribution one can have a process
which is P-divisible, but not CP-divisible.

\subsubsection{Measures of non-Markovianity}

Also for this model we can obtain the explicit expression for the measures of
non-Markovianity according to {\cite{Breuer2009b}} and {\cite{Rivas2010a}}.
The trace distance now reads
\begin{equation}
  D \left( \rho^1 \left( t \right), \rho^2 \left( t \right) \right)  = 
  \sqrt{q^2 \left( t \right) \Delta_p \left( 0 \right)^2 + \left| \Delta_c
  \left( 0 \right) \right|^2 g^2 \left( t \right)},  \label{eq:td2}
\end{equation}
where we have used the same notation as in Eq.~(\ref{eq:pop}) and
Eq.~(\ref{eq:coh}), so that the derivative is
\begin{equation}
  \sigma \left( t, \rho^{1, 2} \left( 0 \right) \right)  =  \frac{\Delta_p
  \left( 0 \right)^2 q \left( t \right) \dot{q} \left( t \right) - \left|
  \Delta_c \left( 0 \right) \right|^2 g \left( t \right) f \left( t
  \right)}{\sqrt{q^2 \left( t \right) \Delta_p \left( 0 \right)^2 + \left|
  \Delta_c \left( 0 \right) \right|^2 g^2 \left( t \right)}}  \label{eq:s2}
\end{equation}
and can grow in the region $\Omega_+$ where $q \left( t \right)$ and $\dot{q}
\left( t \right)$ have the same sign. In this region $\left| q \left( t
\right) \right|$ does increase. The growth is maximal for $\Delta_c
\left( 0 \right) = 0$ and $\Delta_p \left( 0 \right) = 1$, so that the couple
of states which maximize it is given by the projectors on ground and excited
state. As a result, similarly as before we have for the measure of
non-Markovianity introduced by Breuer, Laine and Piilo
\begin{eqnarray}
  \mathcal{N} \left( \Phi \right) & = & \int_{\Omega_+} \mathd t
  \frac{\mathd}{\mathd t}  \left| q \left( t \right) \right| 
  \label{eq:mblp2}\\
  & = & \sum_i \left( \left| q \left( b_i \right) \right| - \left| q \left(
  a_i \right) \right| \right) . \nonumber
\end{eqnarray}
This result for the choice of CPT map $\mathcal{E}_{\pm}$ is right the same as
for the CPT map $\mathcal{E}_z$. This measure becomes nonzero if and only if
P-divisibility is broken. It can be shown that this is generally the case for
a bistochastic map $\Phi$ in $\mathbbm{C}^2$.

The criterion of Rivas, Huelga and Plenio instead assigns to the map a nonzero
measure whenever one of the coefficients $\gamma \left( t \right)$ or $\delta
\left( t \right)$ take on negative values, so that CP-divisibility is broken.
Since $h \left( t \right)$ is always positive, these two functions can take on
negative values only on separate time intervals, as can also be seen from
Fig.~\ref{fig:rates1}. The measure is then given by Eq.~(\ref{eq:defrhp1}),
where according to Eq.~(\ref{eq:defrhp2}) we have $\mathfrak{g} \left( t
\right) = 0$ whenever both $\gamma \left( t \right)$ and $\delta \left( t
\right)$ are positive, while $\mathfrak{g} \left( t \right) = - 2 \gamma
\left( t \right)$ whenever $\gamma \left( t \right)$ is negative, and
$\mathfrak{g} \left( t \right) = - 2 \delta \left( t \right)$ in the
complementary time intervals in which $\delta \left( t \right)$ takes on
negative values. Note that this measure $\mathcal{I} \left( \Phi \right)$ can
become positive even if the measure $\mathcal{N} \left( \Phi \right)$ is zero.
Indeed the latter measure for this dynamics is related to P-divisibility
rather than CP-divisibility.

\subsubsection{Population dynamics}

For the dynamics described by Eq.~(\ref{eq:ms}), with the CPT map given by
$\mathcal{E}_{\pm}$ as in Eq.~(\ref{eq:epm}), coherences and populations
decouple, and the populations obey the same equation as the transition
probabilities of the classical semi-Markov processes considered in
Sect.~\ref{sec:cnm} for $\pi = 1$ and arbitrary waiting time distribution.
This is immediately seen identifying the two components of the probability
vector with the populations in excited and ground state. Setting $P_+ \left( t
\right) = \langle + | \rho \left( t \right) | + \rangle$ and $P_- \left( t
\right) = \langle - \rho \left( t \right) | - \rangle$ one has in fact from
Eq.~(\ref{eq:ms}) with $\mathcal{E}_{\pm}$ the integrodifferential rate
equations
\begin{eqnarray}
  \frac{\mathd}{\tmop{dt}} P_{\pm} \left( t \right) & = & \int^t_0 d \tau k
  \left( t - \tau \right) \left[ P_{\mp} \left( \tau \right) - P_{\pm} \left(
  \tau \right) \right]  \label{eq:fkq}
\end{eqnarray}
corresponding to Eq.~(\ref{eq:fk}) for
\begin{eqnarray}
  W \left( \tau \right) & = & \left(\begin{array}{cc}
    - 1 & 1\\
    1 & - 1
  \end{array}\right) k \left( \tau \right) .  \label{eq:qmatrix}
\end{eqnarray}
Also in the classical case the integrodifferential time evolution equation can
be generally recast in time local form {\cite{Hanggi1978a}}. In this case
Eq.~(\ref{eq:fkq}) would then simply correspond to the diagonal matrix
elements of Eq.~(\ref{eq:tclpm}).

The Kolmogorov distance as in Eq.~(\ref{eq:dkq}) is given by
\begin{eqnarray}
  D_K \left( \left\{ P_+^1 \left( t \right), P_-^1 \left( t \right) \right\},
  \left\{ P_+^2 \left( t \right), P_-^2 \left( t \right) \right\} \right) & =
  & \left| \Delta_p \left( 0 \right) \right| \left| q \left( t \right)
  \right|, \nonumber
\end{eqnarray}
so that $\mathcal{N} \left( \Phi \right)$, being obtained by considering as
initial states the projections onto ground and excited state, is also given by
taking the maximum over the possible initial classical states of the integral
of the derivative $\sigma_K$ of the Kolmogorov distance in the time intervals
in which it is positive. Growth of the Kolmogorov distance again depends on
the behavior of $\left| q \left( t \right) \right|$ only, which determines
whether the map is P-divisible or not. In view of this connection it appears
that one can have non-Markovianity measure $\mathcal{N} \left( \Phi \right)$
for the quantum map $\Phi$ equal to zero even if the dynamics for the
populations can be related to a non-Markovian classical process. Again this is
not too surprising, since the one-point probabilities cannot really keep track
of Markovianity or non-Markovianity in the classical sense, even though in the
non-Markovian case they can show up different behaviors than those typical of
the Markovian one.

\subsubsection{Explicit examples}

At variance with the case of pure dephasing, the two measures of
non-Markovianity do not agree for this model. The measure $\mathcal{N} \left(
\Phi \right)$ becomes positive as soon as P-divisibility is broken, which
depends on the sign of $\gamma \left( t \right)$ only, while $\mathcal{I}
\left( \Phi \right)$ becomes positive even when only CP-divisibility does not
hold, which also depends on the sign of the function $\delta \left( t \right)$
appearing in front of the purely quantum channel $\mathcal{L}_z$, which
determines the dynamics of the coherences. To consider the behavior of the
measures for this model we thus have to consider also the quantity $\delta
\left( t \right)$, which is simply equal to zero for an exponential waiting
time distribution, so that in the proper Markovian case this pure quantum
channel is not available.

For the case of the convolution of two equal exponential distributions
exploiting Eq.~(\ref{eq:fsed}) and \ Eq.~(\ref{eq:gsed}) together with
Eq.~(\ref{eq:gtz}) we have
\begin{eqnarray}
  \delta \left( t \right) & = & \frac{\lambda}{2}  \left( \frac{\lambda t}{1 +
  \lambda t} - \frac{1}{1 + \tmop{cotg} \left( \lambda t \right)} \right), 
  \label{eq:hg1}
\end{eqnarray}
so that both $\gamma \left( t \right)$ and $\delta \left( t \right)$ oscillate
in sign and diverge when $\tmop{cotg} \left( \lambda t \right)$ takes on the
value minus one, as shown in Fig.~\ref{fig:rates1}. In this case both measures
are positive, while considering the convolution of two different exponential
distributions one has thanks to Eq.~(\ref{eq:fged}), Eq.~(\ref{eq:gged}) and
Eq.~(\ref{eq:gtz2})

\begin{eqnarray}
  \delta \left( t \right) & = & \frac{p}{s}  \left( \frac{1}{1 + \xi
  \tmop{Coth} \left( \frac{st}{2} \xi \right)} - \frac{1}{1 + \chi \tmop{Coth}
  \left( \frac{st}{2} \chi \right)} \right) .  \label{eq:hg2}
\end{eqnarray}
If the ratio $\lambda_1 / \lambda_2$ is far enough from one $\gamma \left( t
\right)$ given by Eq.~(\ref{eq:gtz2}) as discussed above stays always
positive, so that one has P-divisibility and the measure $\mathcal{N} \left(
\Phi \right)$ is equal to zero. On the contrary the function $\delta \left( t
\right)$ is negative, so that the coefficient in front of the quantum channel
is always negative and CP-divisibility is violated , thus determining a
positive measure $\mathcal{I} \left( \Phi \right)$. This situation is
considered in Fig.~\ref{fig:rates2}.

As a last example we consider a convex mixture of exponential distributions,
leading to Eq.~(\ref{eq:gtz3}) as well as
\begin{eqnarray}
  h \left( t \right) & = & \frac{\mu \lambda_1 \mathe^{- \lambda_1 t} + \left(
  1 - \mu \right) \lambda_2 \mathe^{- \lambda_2 t}}{\mu \mathe^{- \lambda_1 t}
  + \left( 1 - \mu \right) \mathe^{- \lambda_2 t}} .  \label{eq:h3}
\end{eqnarray}
For this case, independently of the value of the mixing parameter
$\mu$, one has that both $\gamma \left( t \right)$ and $\delta \left(
   t \right)$ stay positive. A dynamics of this kind for an open
quantum system is often termed time-dependent Markovian
{\cite{Laine2010a}}, since at any time the generator is in Lindblad
form. Once again both measures $\mathcal{N} \left( \Phi \right)$ and
$\mathcal{I} \left( \Phi \right)$ give a zero value of
non-Markovianity, despite the fact that the underlying waiting time
distribution is not memoryless, corresponding to a population dynamics
described by a classical semi-Markov process which is not Markovian.

\section{Conclusions and outlook\label{sec:ceo}}

In this paper we have analyzed the notion of non-Markovianity for the dynamics
of open quantum systems, starting from the classical setting and focusing on
concrete examples. While knowledge of a non-Markovian classical process
requires information on all the conditional probability densities, when
studying the dynamics of an open system one only considers the evolution of
the state, expressed by a probability vector in the classical case and a
statistical operator for a quantum system. The notion of non-Markovianity for
classical processes and for dynamical evolutions are therefore by necessity
distinct concepts. One is then naturally lead to the question whether and how
the non-Markovianity of a process reflects itself in the behavior of the
one-point probability. It turns out that the notion of P-divisibility, in the
sense of the existence of well-defined stochastic matrices connecting
probability vectors at different time points, as well as monotonic decrease in
time of the Kolmogorov distance among states arising from different initial
conditions, are typical features of Markovian processes in the classical case.
The lack of these properties can thus be interpreted as signature of
non-Markovianity, and used to quantify it. Note that due to the fact that the
classical definition of non-Markovianity actually involves all $n$-point
probability densities, these signatures as expected indeed provide a different
notion of non-Markovianity, which only gives a sufficient condition in order
to assess non-Markovianity in the original sense. This behavior is shown by
means of example relying on the study of certain semi-Markov processes.

Such signatures of non-Markovianity can be brought over to the quantum
framework, considering the notion of CP-divisibility, corresponding to the
fact that the quantum time evolution map can be arbitrary split into CPT maps
corresponding to intermediate time intervals, and of the trace distance, which
is the quantum version of the Kolmogorov distance. These two criteria are at
the basis of two recently introduced measures of non-Markovianity for open
quantum systems {\cite{Breuer2009b,Rivas2010a}}, which we here compare
considering a quantum counterpart of classical semi-Markov processes.
Moreover, analyzing both the integrodifferential and the time local form of
the equations of motion, we show that one can point to possible signatures of
non-Markovianity to be read directly at the level of the equation, despite the
fact that Markovianity or non-Markovianity itself is a feature of the
solution. In this respect it appears that the time local form of the
equations, despite isolated singularities, is certainly more convenient.

More specifically, the classes of examples discussed in the paper, besides
clarifying the relationship between the distinct notions of non-Markovianity
used for classical processes and for dynamical evolutions, allow an exact
evaluation of the measure of non-Markovianity introduced for quantum dynamical
maps, see e.g. the remarkable result given by Eq.~(\ref{eq:exact}). These
physical examples further allow an explicit comparison of the two measures for
whole classes of quantum dynamical maps. In particular, as shown by
Eq.~(\ref{eq:mblp}) and Eq.~(\ref{eq:mrhp}) and discussed thoroughly in
Sect.~\ref{sec:ee2}, we put into evidence that the measure based on
CP-divisibility {\cite{Rivas2010a}} gives the same infinite value to quite
different time evolutions, at variance with the measure based on the dynamics
of the trace distance {\cite{Breuer2009b}}, which assigns to them different
weights.

In this paper we have discussed the different definitions of non-Markovianity
relevant for classical stochastic processes and dynamical evolutions. The
latter are based on the divisibility of the time evolution map and on the
dynamic behavior of the distinguishability among different initial states.
While these definitions of non-Markovianity can be considered both in the
classical and the quantum case, as discussed in Sect.~\ref{sec:cnm} and
Sect.~\ref{sec:qnm} the original definition of non-Markovianity for a
classical process cannot be transferred as such to the quantum realm, because of basic
principles of quantum mechanics. On the one hand to make statements about the
value of a certain observables at different times, a measurement scheme has
to be specified, which affects the subsequent time evolution; on the other
hand the statistical operator of a quantum systems provide different, and
generally incompatible, classical probability densities for different
observables, as a typical feature of quantum probability with respect to
classical probability {\cite{Holevo2001,Strocchi2005}}. In this respect one
should be satisfied with different definitions and measures of
non-Markovianity within the classical and the quantum framework.

It is clear that physical systems can provide us with much more complicated
dynamics than the one considered in this paper and the recent literature.
However the analysis of these situations sheds light on the basic ideas used
in defining Markovianity or non-Markovianity of a time evolution, and to their
connection with the classical definition of non-Markovian, thus possibly
opening the way for a more refined and general analysis.

\section{Acknowledgments}

BV and AS gratefully acknowledge financial support by MIUR under PRIN2008,
EML by the Graduate School of Modern Optics and Photonics, JP by the Magnus Ehrnrooth Foundation, and HPB by the German Academic Exchange Service.


\begin{thebibliography}{10}
  \bibitem[1]{Breuer2002}H.-P. Breuer and F.~Petruccione, \tmtextit{The Theory
  of Open Quantum Systems} (Oxford University Press, Oxford, 2002)
  
  \bibitem[2]{Stockburger2002a}J.~T. Stockburger and H.~Grabert, Phys. Rev.
  Lett. \tmtextbf{88}, 170407 (2002)
  
  \bibitem[3]{Daffer2004a}S.~Daffer, K.~W\'odkiewicz, J.~D. Cresser, and J.~K.
  McIver, Phys. Rev.~A \tmtextbf{70}, 010304 (2004)
  
  \bibitem[4]{Piilo2008a}J.~Piilo, S.~Maniscalco, K.~H\"ark\"onen, and K.-A.
  Suominen, Phys. Rev. Lett. \tmtextbf{100}, 180402 (2008)
  
  \bibitem[5]{Breuer2008a}H.-P. Breuer and B.~Vacchini, Phys. Rev. Lett.
  \tmtextbf{101}, 140402 (2008)
  
  \bibitem[6]{Shabani2009a}A.~Shabani and D.~A. Lidar, Phys. Rev. Lett.
  \tmtextbf{102}, 100402 (2009)
  
  \bibitem[7]{Chruscinski2010a}D.~Chruscinski and A.~Kossakowski, Phys. Rev.
  Lett. \tmtextbf{104}, 070406 (2010)
  
  \bibitem[8]{Barchielli2010a}A.~Barchielli, C.~Pellegrini, and
  F.~Petruccione, EPL \tmtextbf{91}, 24001 (2010)
  
  \bibitem[9]{Laine2010c}E.-M. Laine, J.~Piilo, and H.-P. Breuer, EPL
  \tmtextbf{92}, 60010 (2010)
  
  \bibitem[10]{Budini2004a}A.~A. Budini, Phys. Rev.~A \tmtextbf{69}, 042107
  (2004)
  
  \bibitem[11]{Wolf2008a}M.~M. Wolf, J.~Eisert, T.~S. Cubitt, and J.~I. Cirac,
  Phys. Rev. Lett. \tmtextbf{101}, 150402 (2008)
  
  \bibitem[12]{Breuer2009b}H.-P. Breuer, E.-M. Laine, and J.~Piilo, Phys. Rev.
  Lett. \tmtextbf{103}, 210401 (2009)
  
  \bibitem[13]{Rivas2010a}A.~Rivas, S.~F. Huelga, and M.~B. Plenio, Phys. Rev.
  Lett. \tmtextbf{105}, 050403 (2010)
  
  \bibitem[14]{Lindblad1976a}G.~Lindblad, Comm. Math. Phys. \tmtextbf{48}, 119
  (1976)
  
  \bibitem[15]{Gorini1976a}V.~Gorini, A.~Kossakowski, and E.~C.~G. Sudarshan,
  J.~Math. Phys. \tmtextbf{17}, 821 (1976)
  
  \bibitem[16]{Laine2010a}E.-M. Laine, J.~Piilo, and H.-P. Breuer, Phys.
  Rev.~A \tmtextbf{81}, 062115 (2010)
  
  \bibitem[17]{Hayashi2006}M.~Hayashi, \tmtextit{Quantum Information}
  (Springer-Verlag, Berlin, 2006)
  
  \bibitem[18]{Wolf2008b}M.~M. Wolf and J.~I. Cirac, Commun. Math. Phys.
  \tmtextbf{279}, 147 (2008)
  
  \bibitem[19]{Norris1999}J.~R. Norris, \tmtextit{Markov Chains} (Cambridge
  University Press, Cambridge, 1999)
  
  \bibitem[20]{Chruscinski2010c}D.~Chruscinski, A.~Kossakowski, P.~Aniello,
  G.~Marmo, and F.~Ventriglia, Open Syst. Inf. Dyn. \tmtextbf{17}, 255 (2010)
  
  \bibitem[21]{Chruscinski-xxx2}D.~Chruscinski and A.~Kossakowski, e-print
  arXiv:1010:4745v1 (2010)
  
  \bibitem[22]{Chruscinski-xxx3}D.~Chruscinski, A.~Kossakowski, and A.~Rivas,
  e-print arXiv:1102:4318v1 (2011)
  
  \bibitem[23]{Rivas-xxx}A.~Rivas and S.~F. Huelga, e-print arXiv:1104:5242v1
  (2011)
  
  \bibitem[24]{Karimipour-xxx}V.~Karimipour and L.~Memarzadeh, e-print
  arXiv:1105:2728v1 (2011)
  
  \bibitem[25]{VanKampen1998a}N.~van Kampen, Braz. J. Phys. \tmtextbf{28}, 90
  (1998)
  
  \bibitem[26]{Gillespie1998a}D.~T. Gillespie, Am. J. Phys. \tmtextbf{66}, 533
  (1998)
  
  \bibitem[27]{Hanggi1977a}P.~H\"anggi and H.~Thomas, Z. Phys. B
  \tmtextbf{26}, 85 (1977)
  
  \bibitem[28]{Hanggi1982a}P.~H\"anggi and H.~Thomas, Phys. Rep.
  \tmtextbf{88}, 207 (1982)
  
  \bibitem[29]{Feller1971}W.~Feller, \tmtextit{An introduction to probability
  theory and its applications. Vol. II} (John Wiley \& Sons Inc., New York,
  1971)
  
  \bibitem[31]{Feller1964a}W.~Feller, Proc. Nat. Acad. Sci. U.S.A.
  \tmtextbf{51}, 653 (1964)
  
  \bibitem[32]{Cox1965}D.~R. Cox and H.~D. Miller, \tmtextit{The theory of
  stochastic processes} (John Wiley \& Sons Inc., New York, 1965)
  
  \bibitem[33]{Medhi1994}J.~Medhi, \tmtextit{Stochastic Processes} (John Wiley
  \& Sons, 1994)
  
  \bibitem[34]{Fuchs1999a}C.~A. Fuchs and J.~van de Graaf, IEEE Trans. Inf.
  Th. \tmtextbf{45}, 1216 (1999)
  
  \bibitem[35]{Nielsen2000}M.~Nielsen and I.~Chuang, \tmtextit{Quantum
  Computation and Quantum Information} (Cambridge University Press, Cambridge,
  2000)

   \bibitem[30]{Breuer2009a}H.-P. Breuer and B.~Vacchini, Phys. Rev.~E
  \tmtextbf{79}, 041147 (2009)
   
  \bibitem[36]{Smirne2010b}A.~Smirne and B.~Vacchini, Phys. Rev.~A
  \tmtextbf{82}, 022110 (2010)
  
  \bibitem[37]{Choi1975a}M.~D. Choi, Lin. Alg. Appl. \tmtextbf{10}, 285 (1975)
  
  \bibitem[38]{Haikka2011a}P.~Haikka, J.~D. Cresser, and S.~Maniscalco, Phys.
  Rev.~A \tmtextbf{83}, 012112 (2011)
  
  \bibitem[39]{Ross2007}S.~M. Ross, \tmtextit{Introduction to probability
  models} (Academic Press, Burlington, MA, 2007)
  
  \bibitem[40]{Hanggi1978a}P.~H\"anggi, H.~Thomas, H.~Grabert, and P.~Talkner,
  J.~Stat. Phys. \tmtextbf{18}, 155 (1978)
  
  \bibitem[41]{Holevo2001}A.~S. Holevo, \tmtextit{Statistical Structure of
  Quantum Theory}, Vol. m 67 of \tmtextit{Lecture Notes in Physics} (Springer,
  Berlin, 2001)
  
  \bibitem[42]{Strocchi2005}F.~Strocchi, \tmtextit{An introduction to the
  mathematical structure of quantum mechanics} (World Scientific, 2005)
\end{thebibliography}
\end{document}